\documentclass[aps,prd,twocolumn,amsmath,amssymb,floatfix,nofootinbib,10pt]{revtex4}

\usepackage{graphicx,epsfig}
\usepackage{amsmath,amssymb,color,hyperref}
\definecolor{linkcolor}{rgb}{0,0,0.25}
\hypersetup{
  colorlinks=true,        % false: boxed links; true: colored links
  linkcolor=linkcolor,    % color of internal links
  citecolor=linkcolor,    % color of links to bibliography
  filecolor=linkcolor,    % color of file links
  urlcolor=linkcolor      % color of external links
}

\newcommand{\eg}{e.g.}
\newcommand{\ie}{i.e.}
\newcommand{\Fermi}{\emph{Fermi}}
\newcommand{\VL}{Via Lactea}
\newcommand{\aquarius}{Aquarius}
\newcommand{\CDM}{CDM}
\newcommand{\LCDM}{\ensuremath{\Lambda}\CDM}
\newcommand{\NFW}{NFW}
\newcommand{\GNFW}{G\NFW}
\newcommand{\DM}{DM}
\newcommand{\SM}{SM}

\newcommand{\somm}{\ensuremath{S}}
\newcommand{\mdm}{\ensuremath{m_{\chi}}}
\newcommand{\mv}{\ensuremath{m_{\phi}}}
\newcommand{\dd}{\mathrm{d}}
\newcommand{\Eqnname}{Equation}
\newcommand{\eqnname}{equation}
\newcommand{\Ngamma}{\ensuremath{N_{\gamma}}}
\newcommand{\Ngammai}{\ensuremath{N_{\gamma,i}}}
\newcommand{\Eth}{\ensuremath{E_{\mathrm{th}}}}
\newcommand{\sigmaannv}{\ensuremath{\langle\sigma v\rangle}}
\newcommand{\los}{los}
\newcommand{\order}{\ensuremath{\mathcal{O}}}
\newcommand{\sigv}{\ensuremath{\sigma_v}}
\newcommand{\lum}{\ensuremath{\mathcal{L}}}
\newcommand{\lumsmooth}{\ensuremath{\lum_{\mbox{{\footnotesize sm}}}}}
\newcommand{\lumsmoothc}{\ensuremath{\lum_{\mbox{{\footnotesize sm}},v=c}}}

\newcommand{\lumsub}{\ensuremath{\lum_{\mbox{{\footnotesize sub}}}}}
\newcommand{\boost}{\ensuremath{B}}
\newcommand{\redboost}{\ensuremath{\mathcal{B}_{\mbox{{\footnotesize red}}}}}
\newcommand{\totalboost}{\ensuremath{\mathcal{B}}}
\newcommand{\redr}{\ensuremath{\tilde{r}}}
\newcommand{\rhos}{\ensuremath{\rho_s}}
\newcommand{\rs}{\ensuremath{r_s}}
\newcommand{\dist}{\ensuremath{D}}
\newcommand{\norm}{\ensuremath{\mathcal{N}}}
\newcommand{\normnfw}{\ensuremath{\norm_1}}
\newcommand{\normeinasto}{\ensuremath{\norm_2}}
\newcommand{\rhominustwo}{\ensuremath{\rho_{-2}}}
\newcommand{\rminustwo}{\ensuremath{r_{-2}}}

\newcommand{\alphaEinasto}{\ensuremath{\alpha}}
\newcommand{\Msol}{\ensuremath{M_{\odot}}}
\newcommand{\Msun}{\Msol}
\newcommand{\Mvir}{\ensuremath{M_{\mbox{{\footnotesize vir}}}}}
\newcommand{\Rvir}{\ensuremath{r_{\mbox{{\footnotesize vir}}}}}
\newcommand{\conc}{\ensuremath{c}}
\newcommand{\rhoc}{\ensuremath{\rho_{\mbox{{\footnotesize c}}}}}
\newcommand{\omegam}{\ensuremath{\Omega_{m}}}
\newcommand{\vcirc}{\ensuremath{v_{\mbox{{\footnotesize c}}}}}
\newcommand{\vmax}{\ensuremath{v_{\mbox{{\footnotesize max}}}}}
\newcommand{\rmax}{\ensuremath{r_{\mbox{{\footnotesize max}}}}}
\newcommand{\dSph}{dSph}
\newcommand{\dSphs}{dSphs}
\newcommand{\SDSS}{SDSS}
\newcommand{\degree}{^\circ}
\newcommand{\Jtilde}{\ensuremath{\tilde{J}}}

\def\urltilda{\kern -.15em\lower .7ex\hbox{\~{}}\kern .04em}

\newcommand{\aap}{Astron.~Astrophys.}
\renewcommand{\nat}{Nature}
\newcommand{\jhep}{J.~High.~Energy.~Phys.}
\newcommand{\astropartphys}{Astropart.~Phys.}
\newcommand{\mnras}{Mon.~Not.~Roy.~Astron.~Soc.}
\newcommand{\apjs}{Astrophys.~J.~Suppl.}
\newcommand{\apjl}{\apj}
\newcommand{\aj}{Astron.~J.}
\newcommand{\jcap}{J.~Cosmol.~Astropart.~Phys.}
\newcommand{\araa}{Ann.~Rev.~Astron.~Astrophys.}

\begin{document}

\title{Substructure Boosts to Dark Matter Annihilation from Sommerfeld Enhancement}
\author{Jo Bovy} \affiliation{Center for
Cosmology and Particle Physics, Department of Physics\\ New York
University, New York, NY 10003}

\begin{abstract}
The recently introduced Sommerfeld enhancement of the dark matter
annihilation cross section has important implications for the
detection of dark matter annihilation in subhalos in the Galactic
halo. In addition to the boost to the dark matter annihilation cross
section from the high densities of these subhalos with respect to the
main halo, an additional boost caused by the Sommerfeld enhancement
results from the fact that they are kinematically colder than the
Galactic halo. If we further believe the generic prediction of \CDM\
that in each subhalo there is an abundance of substructure which is
approximately self-similar to that of the Galactic halo, then I show
that additional boosts coming from the density enhancements of these
small substructures and their small velocity dispersions enhance the
dark matter annihilation cross section even further. I find that very
large boost factors ($10^5$ to $10^9$) are obtained in a large class
of models. The implications of these boost factors for the detection
of dark matter annihilation from dwarf Spheroidal galaxies in the
Galactic halo are such that, generically, they outshine the background
gamma-ray flux and are detectable by the Fermi Gamma-ray Space
Telescope.
\end{abstract}

%\keywords{}
\maketitle

\section{Introduction}
In the standard flat, Gaussian, adiabatic, and scale-invariant \LCDM\
paradigm $\sim\!80$ \% of the matter density of the Universe is in the
form of dark matter (\DM) \cite{2008arXiv0803.0547K}. The properties
of dark matter are largely unconstrained, although a weakly
interacting massive particle (WIMP) with a mass $\sim\!100$ GeV -- 1
TeV has many attractive features, including that it leads to a relic
density of dark matter that is remarkably close to the measured value
\cite{1996PhR...267..195J}. No interaction of dark matter either with
itself or with standard model (\SM) particles other than gravity has
ever been observed. However, at the present day we cannot exclude that
the dark sector is brimming with a rich phenomenology at
weak-interaction scales or lower.

One exciting possibility is that dark matter self-annihilates at a
level observable in the Galactic halo today. Detecting this
self-annihilation seems unlikely considering the standard
self-annihilation cross section; $\sigmaannv_{\mathrm{ann}} \lesssim 3
\times 10^{-26} $ cm$^3$ s$^{-1}$ for consistency with the observed
relic density. However, there are various mechanisms to boost the
annihilation signal to a value larger by a few orders of magnitude,
e.g., clumpiness of the dark halo
\cite{1993ApJ...411..439S,1999PhRvD..59d3506B,2008A&A...479..427L,2008ApJ...686..262K}
or a combination of coannihilations and Sommerfeld enhancement
\cite{2005PhRvD..72j3521P,2008arXiv0812.0360L}. The possibility of
detecting the \DM\ annihilation signal has attracted much attention
recently because of the anomalous excesses of high energy electrons
and positrons reported by two different experimental groups: the
Payload for Antimatter Exploration and Light-nuclei Astrophysics
(PAMELA) satellite, which saw a sharp upturn in the positron fraction
$e^+/(e^++e^-)$ from 10 to 100 GeV \cite{Adriani:2008zr}, and the
Advanced Thin Ionization Calorimeter (ATIC) group, which reported an
excess over the expected background in the combined number of
electrons and positrons at energies 500 to 800 GeV
\cite{2008Natur.456..362C}. The PAMELA result on the positron fraction
confirms earlier results
\cite{1969ApJ...158..771F,1975ApJ...199..669B,1987ApJ...312..183M,1994ApJ...436..769G,Barwick:1997ig,Beatty:2004cy,Aguilar:2007yf}
and can be explained by \DM\ annihilation
\cite{2002PhLB..536..263K,2004PhRvD..69j3509H,2008arXiv0810.5344C}. There
is still a good chance that these anomalies can be obviated by
conventional astrophysical effects, e.g., pulsars
\cite{1995A&A...294L..41A,2008arXiv0810.1527H,2008arXiv0810.2784Y,2008arXiv0812.4457P}. Nevertheless,
a plethora of models has been proposed in the last few months
explaining the PAMELA and ATIC results in terms of \DM\ annihilation
into electron-positron pairs (e.g.,
\cite{2008arXiv0810.5557H,ArkaniHamed:2008qn,2008arXiv0811.0399F,2008arXiv0811.1555I,2008arXiv0811.3357C,2008arXiv0811.3641C,2008arXiv0812.2196A}),
a description that also naturally accounts for other experimental
anomalies, such as the ``WMAP haze''
\cite{Finkbeiner:2003im,Dobler:2007wv,Hooper:2007kb}, EGRET
observations of an excess of high-energy gamma-rays in the Galactic
center \cite{Strong:2005zx}, and recent HESS measurements of the flux
of very high-energy electrons \cite{2008arXiv0811.3894H}. Model
building is complicated by the fact that PAMELA did not observe a
similar excess in the ratio of anti-protons to protons
\cite{2008arXiv0810.4994A}, which means that the coupling of the \DM\
particles to quarks must somehow be suppressed.

The largest obstacle to modeling the PAMELA/ATIC signal with \DM\
annihilation is the very large boost factor ($10^4$ to $10^5$) with
respect to the fiducial $\langle \sigma v \rangle_{\mathrm{ann}}
\approx 3 \times 10^{-26} $ cm$^3$ s$^{-1}$ required to produce an
observable annihilation flux. The boost factor from the clumpiness of
the dark matter halo is expected to be small
\cite{2008A&A...479..427L,2008Natur.454..735D,2008ApJ...686..262K,2008Natur.456...73S},
typically of the order 1 to 10. Therefore, a different mechanism must
be responsible for the large boosts necessary. One such mechanism that
has received much attention lately is the Sommerfeld enhancement
\cite{sommerfeld31a,2003PhRvD..67g5014H,2004PhRvL..92c1303H,2005PhRvD..71f3528H,2005PhRvD..71a5007H,2006PhRvD..73e5004H,2008NuPhB.800..204C,2008JHEP...07..058M,2008arXiv0812.0559M,2008arXiv0812.0360L}. The
Sommerfeld enhancement is a generic effect present whenever there is
an attractive force acting between the dark matter particles, \eg, a
Yukawa interaction or a gauge interaction through vector bosons
\cite{ArkaniHamed:2008qn}. The main effect of this attractive force is
to enhance the annihilation cross section $\sigma v$ with a factor
proportional to $\beta^{-1}$ ($\beta \equiv v/c$); this is known as
``$1/v$'' enhancement. There are regions in the parameter space made
up of the dark matter mass \mdm, the force carrier mass \mv, and the
force ``fine-structure constant'' $\alpha \equiv$ coupling$^2$/$4 \pi$
in which the enhancement is near a resonance, where it approximately
behaves as $\beta^{-2}$. However, the enhancement saturates at a
certain $\beta$ because of the finite range of the attractive
force---large ($\alpha \sim\!10^{-1}$ to $10^{-3}$) attractive forces
with infinite range lead to a burst of \DM\ annihilation in the first
\DM\ halos formed at $z$ $\sim$ 100 -- 200 which would lead to
observable effects incompatible with measurements of the diffuse
extragalactic gamma-ray background today and the cosmic microwave
background \cite{2008arXiv0810.3233K}. Since the typical velocities of
dark matter particles in the Galactic halo today are of the order of
the velocity dispersion of the halo, which is non-relativistic
($\beta\sim 10^{-3}$), the cross section of dark matter particles
annihilating in the halo could be significantly enhanced by this
Sommerfeld effect.

Both observations
\cite{1994Natur.370..194I,1999Natur.402...53H,2005ApJ...626L..85W,2007ApJ...654..897B,koposov}
as well as numerical simulations
\cite{1998MNRAS.300..146G,johnston98a,1999ApJ...524L..19M,1999ApJ...522...82K,1999MNRAS.307..495H,2003MNRAS.339..834H,2008Natur.454..735D,2008MNRAS.391.1685S}
have firmly established that dark matter halos in \CDM\ are not smooth
structures, but that they are filled with a large amount of
substructure. These substructures are both in the form of subhalos,
self-bound objects orbiting in the Galactic halo, and in the form of
streams, relics of tidally disrupted subhalos. Since these
substructures are overdense regions in the halo, the dark matter
annihilation cross section will be correspondingly higher:
Annihilation cross sections scale as the density squared. The effects
of these kinds of substructure boosts have been studied extensively
during the last decade
\cite{2004PhRvD..70b3512B,2004PhRvD..69d3501K,2007ApJ...657..262D,2007PhRvD..75h3526S,2008A&A...479..427L,2008MNRAS.384.1627P,2008ApJ...678..614S,2008ApJ...686..262K,2008Natur.456...73S,2008arXiv0809.2781G}. However,
these subhalos are not only denser than the Galactic halo in which
they orbit, they are also kinematically colder than the Galactic halo,
with velocity dispersions that are typically two orders of magnitude
smaller \cite{1998ARA&A..36..435M,2007ApJ...670..313S}. This means
that the Sommerfeld enhancement in these subhalos can be many orders
of magnitude larger than the Sommerfeld enhancement acting in the
Galactic halo taken as a whole.

In addition to this larger boost caused by the small velocity
dispersion in the subhalos, these subhalos themselves are filled with
an abundance of substructure which is approximately self-similar to
that of the main halo. Each of these even smaller substructure has a
larger density than the subhalo in which they are orbiting (quantified
by the well established concentration-mass relationships for \CDM\
halos \cite{2001MNRAS.321..559B,2001ApJ...554..114E}) as well as a
smaller velocity dispersion. This opens up the possibility that the
dark matter annihilation signal in these already cold subhalos can be
boosted even further by contribution of their substructure.

In this paper we quantify the magnitude of this new substructure boost
and discuss its implications for the detection of dark matter
annihilation from subhalos in the Galactic halos. As we will show, the
total combined boost from Sommerfeld enhancement and substructure can
reach very large values (10$^5$ to 10$^9$), and these boosts are only
weakly dependent on the assumptions that we have to make about the
abundance of substructure and the \DM\ density profile. We use these
total boosts together with published \DM\ density profiles of some of
the dwarf Spheroidal satellite galaxies of the Milky Way to predict
the total flux of gamma-rays from \DM\ annihilation that would be
detectable by the Fermi Gamma-ray Space Telescope (\Fermi) and find
that in most of the models that posit the Sommerfeld enhancement the
\DM\ annihilation signal from the dwarf Spheroidals shines above the
diffuse extragalactic gamma-ray background, which is the only
background signal at the high Galactic latitudes of these satellite
galaxies. Finally, we also consider the prospects for detecting
smaller subhalos of the Galaxy by computing annihilation fluxes of
subhalos in the \VL\ II simulation. Whether these subhalos will be
detectable in the near future depends on the exact details of the
Sommerfeld enhancement.

\section{The Sommerfeld enhancement}

The simplest example of the Sommerfeld effect that contains all of the
generic features described in the introduction is that from an
attractive Yukawa type force mediated by a spin-0 boson $\phi$ between
the \DM\ particles, and we adopt that description here. The Sommerfeld
effect is a non-perturbative effect in the sense that in the language
of quantum field theory it comes about from Feynman diagrams in which
the force carrier is exchanged many times before the annihilation
actually happens, giving rise to so called ``ladder
diagrams''. However, we can study the Sommerfeld effect in simple,
non-relativistic quantum mechanics (see the appendix in
Ref.~\cite{ArkaniHamed:2008qn}). We are instructed to solve the radial
Schr{\"o}dinger equation (in natural units)
\begin{equation}\label{eq:sommdiff1}
\frac{1}{\mdm} \frac{\dd^2 \psi(r)}{\dd r^2} + \frac{\alpha}{r}e^{-\mv r} \psi(r) = -\mdm \beta^2 \psi(r)\, ,
\end{equation}
subject to the boundary condition $\dd\psi/\dd r = i \mdm \beta \psi$
as $r \rightarrow \infty$. Here $\mdm$ and $\mv$ are the mass of the
dark matter and the force carrier, respectively. Using the
substitution $r \rightarrow \alpha \mdm r$ this becomes
\begin{equation}\label{eq:sommdiff2}
\frac{\dd^2 \psi(r)}{\dd r^2} + \frac{1}{r}e^{-\mv r/(\alpha \mdm)} \psi(r) = -\frac{\beta^2}{\alpha^2} \psi(r)\, ,
\end{equation}
with boundary condition
\begin{equation}
\psi \propto e^{i\beta r/\alpha} \qquad \mbox{ as } r \rightarrow\infty\, .
\end{equation}
The Sommerfeld enhancement \somm\ is then given by
\begin{equation}
\somm = \frac{|\psi(\infty)|^2}{|\psi(0)|^2}\, .
\end{equation}
\Eqnname~(\ref{eq:sommdiff2}) shows that the Sommerfeld enhancement is
only a function of two variables, which can be chosen as
$\alpha/\beta$ and $\alpha \mdm/\mv$. The qualitative features of
\eqnname\ (\ref{eq:sommdiff2}) have been widely discussed and,
therefore, we will merely mention them here without proof (see, \eg,
Refs.~\cite{ArkaniHamed:2008qn,2008arXiv0812.0360L}). When the kinetic
energy of the collision is much larger than the product of the force
carrier mass and the coupling, then the Yukawa interaction can be
approximated as an attractive Coulomb interaction, for which an
analytical solution in terms of hypergeometric functions exists. The
Sommerfeld enhancement is simply given by
\begin{equation}
\somm = \frac{\pi \alpha}{\beta} (1 - e^{-\pi\alpha/\beta})^{-1}\, .
\end{equation}
From this we see why the Sommerfeld enhancement is often referred to
as a ``$1/v$'' enhancement for small velocities $\beta \ll 1$. In the
limit of a vanishing interaction ($\alpha/\beta \ll 1$) the Sommerfeld
enhancement disappears as $\somm \rightarrow 1$. The $1/v$ behavior,
which diverges for small $\beta$ and leads to an overproduction of
high-energy photons in the early Universe \cite{2008arXiv0810.3233K},
saturates as a consequence of the finite range of the Yukawa
interaction at a velocity $\beta \approx \mv/\mdm$ and settles on a
value $\somm \approx \alpha \mdm/\mv$. However, at some parameter
combinations the Yukawa potential has bound states, at which there are
large boosts to the annihilation cross section which go approximately
as $\somm \propto \beta^{-2}$. These resonance regions, however, are
cut-off at small $\beta$ by the finite lifetime of the bound state. We
expect these three different behaviors ($1/v$, saturation at low $v$,
and resonance effects from the existence of bound states) to be
generic among different models of the Sommerfeld enhancement (see also
the discussion in the appendix of Ref.~\cite{ArkaniHamed:2008qn}).

\begin{figure}[t]
\centering
\includegraphics{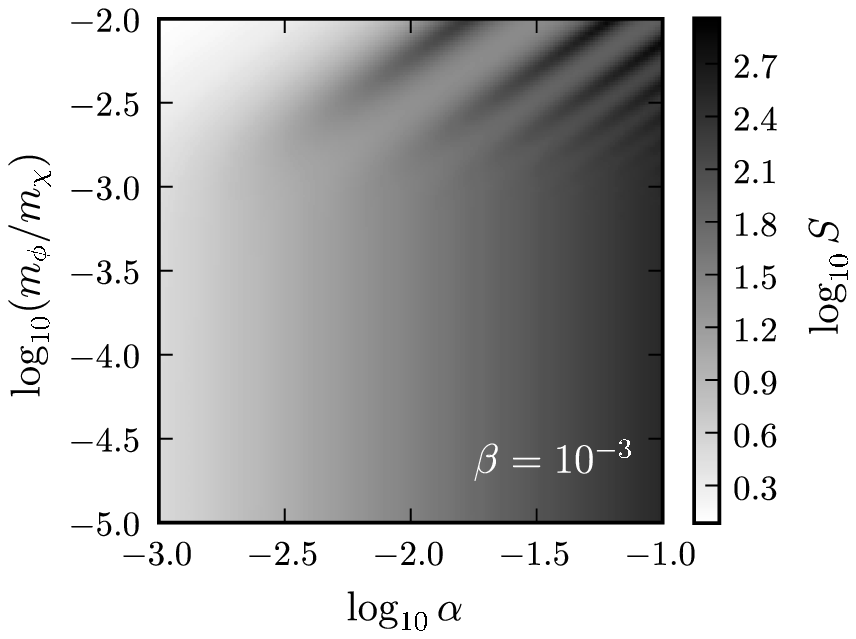}
\\
\includegraphics{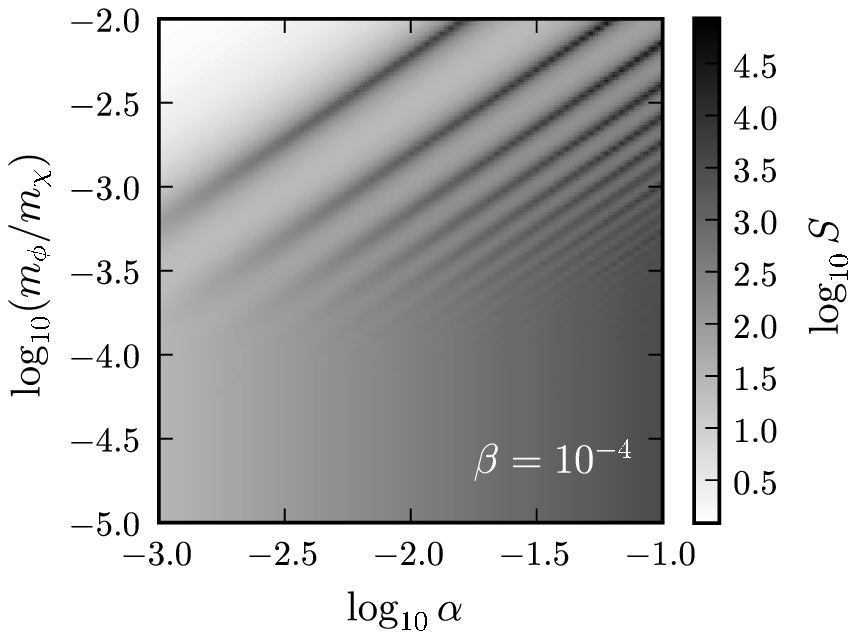}
\caption{Sommerfeld enhancement as a function of the particle physics
parameters for two different values of the relative velocity
$\beta$. Because of the scaling properties of \eqnname\
(\ref{eq:sommdiff2}) the behavior of the Sommerfeld enhancement as a
function of $\log_{10}\beta$ can actually be read off along diagonals
in the plane.}%\
\label{fig:sommb34}%
\end{figure}

\Eqnname\ (\ref{eq:sommdiff2}) can also be numerically integrated
efficiently, by integrating the solution $\psi(r) = e^{i\beta
r/\alpha}$ inwards from a radius satisfying
\begin{equation}
\frac{1}{r}e^{-\mv r/(\alpha \mdm)} \ll \frac{\beta^2}{\alpha^2} \, .
\end{equation}
In the Coulomb regime ($\mdm \beta^2 \gg \alpha \mv$) this means that
the initial value for the integration should be such that $r \gg
\alpha^2/\beta^2$; otherwise we should start at $r \gg \alpha
\mdm/\mv$. Figure \ref{fig:sommb34} shows the Sommerfeld enhancement
in the parameter plane made of $\alpha$ and $\mv/\mdm$. All of the
qualitative features we discussed above are present: In the Coulomb
regime (lower part of the two panels) we see that the enhancement
scales as $\alpha$, saturating at $\somm \approx 1$ when $\alpha$
becomes small; The fact that the upper left corner of both panels is
approximately the same shows that the Sommerfeld enhancement saturates
at $\beta \approx \mv/\mdm$; The sharp diagonal lines in the upper
right of both panels show the resonance region, in which the
enhancement is sharply peaked. Because of the scaling properties of
\eqnname\ (\ref{eq:sommdiff2}), shifting $\log\beta$ is equivalent to
shifting both $\log\alpha$ and $\log \mv/\mdm$ by the same
amount. Therefore, the $\beta$ behavior of the enhancement can also be
read off from these figures along diagonals. Inspection of the
resonance diagonals in the second panel of Figure \ref{fig:sommb34}
shows that the enhancement grows as $\beta^{-2}$ in these regions.

Of course, when considering the Sommerfeld enhancement of a halo of a
certain velocity dispersion $\sigv$ we need to average the Sommerfeld
enhancement over the distribution of relative velocities of that
halo. We will approximate the one-particle velocity distribution of
the halo as a single truncated Maxwell-Boltzmann distribution with
velocity dispersion $\sigv$
\begin{equation}\label{eq:maxboltz}
f(v) \propto \left\{
\begin{array}{rl}
v^2\, e^{-v^2/2\sigv^2}  & \qquad v \leq v_{\mathrm{esc}}\\
0 & \qquad v > v_{\mathrm{esc}}\, .\\
\end{array} \right.
\end{equation}
This approximation of a single Maxwell-Boltzmann
distribution for the velocity distribution means that we assume that
the halo has a constant velocity dispersion over the halo, \ie, that
the halo is isothermal. This is not necessarily the case---indeed, it
is not if we believe that the density profiles of \DM\ halos are well
fit by Einasto or \NFW\ profiles---but it is not a bad assumption for
intermediate distances from the halo's center, \ie, between the
innermost and outermost parts of the halo
\cite{2007ApJ...667L..53W}. A more detailed treatment of the
Sommerfeld enhancement in subhalos should take the non-uniform
velocity dispersion in subhalos into account. Note that in the pure
$1/v$ regime the averaging will tend to sustain the $1/v$ behavior up
to a factor $\order(1)$, as we have that
\begin{equation}
\int_0^\infty \dd v \, v^2\, e^{-v^2/4\sigv^2}
\somm(v)\left/\int_0^\infty \dd v \, v^2\, e^{-v^2/4\sigv^2} \right.\propto
\frac{1}{\sigv}\, ,
\end{equation}
since the truncation does not significantly influence the result of
these integrals. Similarly, in the resonance region the $1/v^2$
behavior is conserved by the averaging. When the enhancement has
leveled off the averaging has no effect.

\begin{figure*}[t]
\includegraphics[clip=]{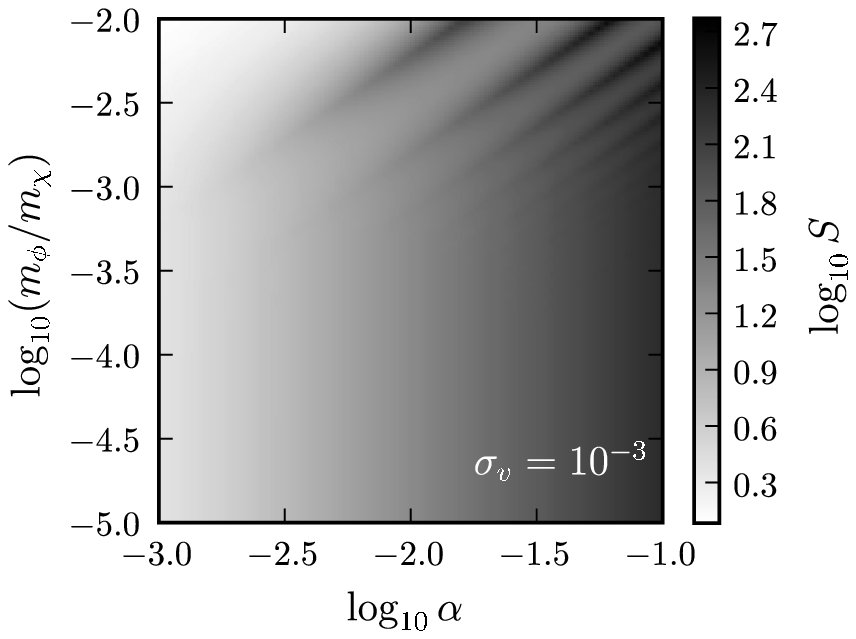}%%BoundingBox: 178 291 400 474
\includegraphics[clip=]{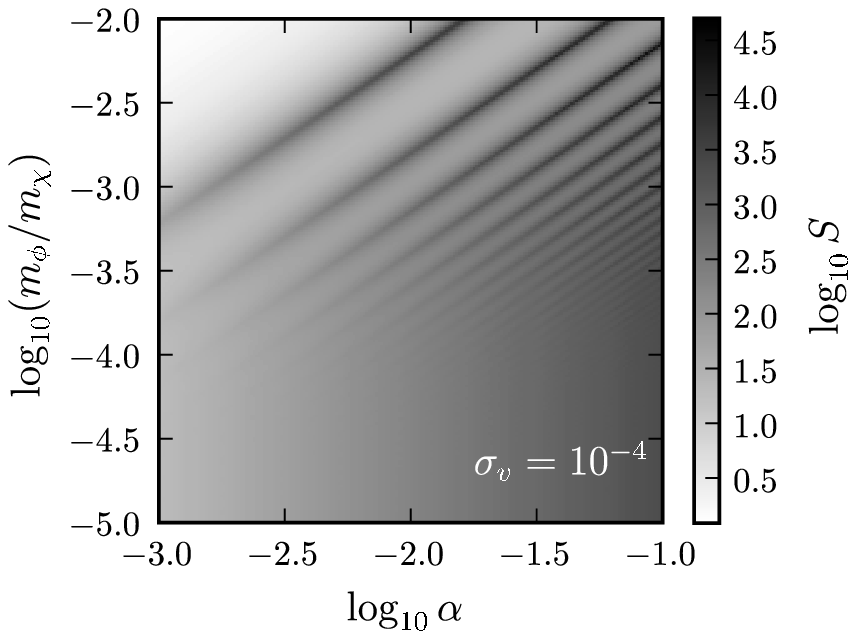}\\%%BoundingBox: 190 291 418 474
\includegraphics[clip=]{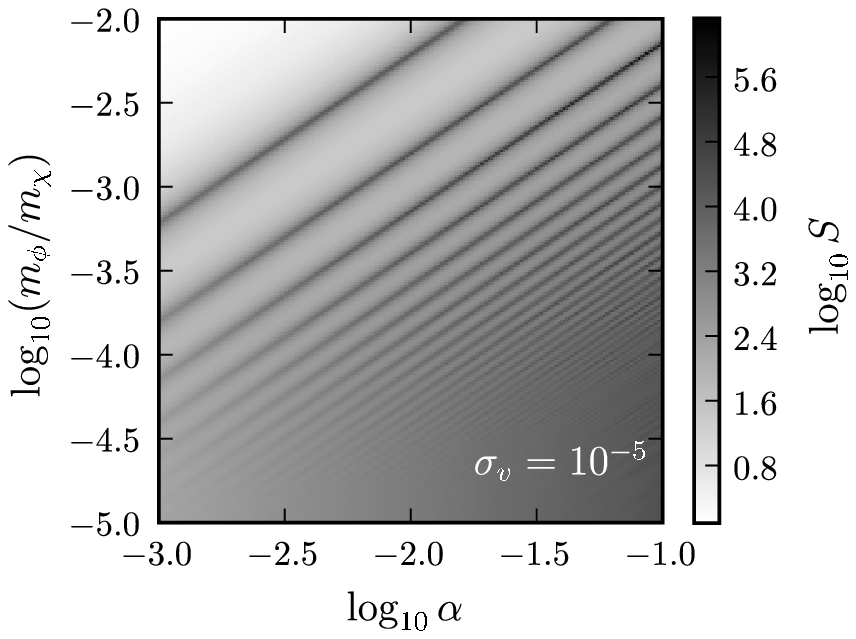}%%BoundingBox: 178 291 400 474
\includegraphics[clip=]{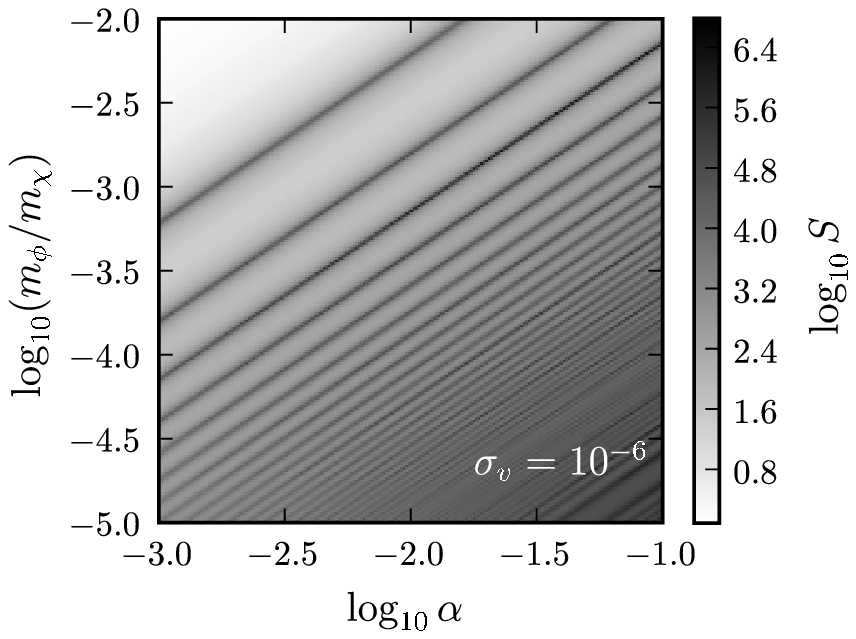}%%BoundingBox: 190 291 418 474
\caption{Sommerfeld enhancement factor as a function of the particle
physics parameters assuming a one-particle Maxwell-Boltzmann velocity
distribution with velocity dispersion \sigv\ for four different orders
of magnitude of the velocity dispersion (all velocities are relative
to the speed of light).}%
\label{fig:sommavg}%
\end{figure*}

In Figure \ref{fig:sommavg} we show this averaged Sommerfeld
enhancement for four different decades of velocity dispersion. The
overall behavior of the Sommerfeld enhancement seen in Figure
\ref{fig:sommb34} is indeed conserved well by the averaging: this has
to do with the fact that the most probable relative velocity in a
Maxwell-Boltzmann relative velocity distribution is the velocity
dispersion. The Sommerfeld enhancement attains very large values for
subhalos with very small velocity dispersions, mostly, but not
exclusively, in the resonance regions. This means that the smallest
subhalos that exist in \CDM\ will get large boosts to their \DM\
annihilation from Sommerfeld enhancement.

\section{Substructure boosts from Sommerfeld enhancement}
The photon flux from \DM\ annihilation from a solid angle
$\Delta\Omega$ along a given line-of-sight (\los) is given by
\begin{equation}\label{eq:flux}
\begin{split}
\frac{\dd \Ngamma}{\dd A \,\dd t}  = &
\int_{\Eth}^{\mdm}\dd E\, \sum_i \frac{\dd
\Ngammai}{\dd E} \,\frac{\sigmaannv_i}{\mdm^2}\,
\somm\left(\mv/\mdm,\alpha,\sigv\right)\,\\
& \times \int_{\Delta\Omega}\,\frac{\dd\Omega}{4\pi}\,\int_{\mathrm{\los}} \dd l\,
\rho^2(l)\, ,
\end{split}
\end{equation}
where \Eth\ is the threshold energy of the detector, $i$ denotes
(possibly) different final states of the \DM\ annihilation, $\frac{\dd
\Ngammai}{\dd E}$ is the spectrum of the annihilation to that state,
$\sigmaannv_i$ is the annihilation cross section to state $i$ (without
the Sommerfeld enhancement), $\somm\left(\mv/\mdm,\alpha,\beta\right)$
is the Sommerfeld enhancement which we have factored out of the cross
section, and $\rho(l)$ is the \DM\ density along the line-of-sight. We
emphasize that this framework also allows us to consider different
final annihilation products, such as the electrons and positrons
observed by PAMELA/ATIC.

We will focus our analysis on the ``structure quantity'' \lum($M$),
which we define here as
\begin{equation}\label{eq:structquant}
\lum(M)\equiv\somm\left(\mv/\mdm,\alpha,\sigv\right)\,\int_{\Delta\Omega}\,\frac{\dd\Omega}{4\pi}\int_{\mathrm{\los}}
\dd l\, \rho^2(l)
\end{equation}
This structure quantity does not only depend on the mass and internal
properties of the \DM\ halos in which we are interested, it depends on
the particle physics parameters as well. We will keep all of the other
factors in \eqnname\ (\ref{eq:flux}) fixed. However, in order to
derive actual numbers for the photon flux from \DM\ annihilation, we
briefly describe the other quantities appearing in \eqnname\
(\ref{eq:flux}).

\subsection{Some details of the particle physics model}\label{sec:partphys}

For the annihilation cross section prior to any Sommerfeld enhancement
we will adopt the fiducial value $\sigmaannv = 3 \times 10^{-26} $
cm$^3$ s$^{-1}$, and we choose a fiducial mass for the \DM\ particle
of \mdm\ = 700 GeV. The energy spectrum of the \DM\ annihilation
depends on the model used to describe the \DM\ and its interactions;
it is outside of the scope of this paper to provide a detailed
treatment of this for different \DM\ models. Therefore, we will adopt
an annihilation spectrum for \DM\ annihilation into photons
appropriate for the newer models that were proposed to explain the
PAMELA/ATIC results. Specifically, we use a gamma ray spectrum from a
cascade annihilation model. In these models the \DM\ annihilates into
a new light degree of freedom which then in turn decays into leptons
and photons. The injection spectrum per \DM\ annihilation is given by
\cite{2009arXiv0901.2926M}
\begin{equation}\label{eq:annspectrum}
\begin{split}
\frac{\dd \Ngamma}{\dd E} = &\frac{\alpha_{\mbox{{\footnotesize EM}}}}{\pi} \frac{1+(1-E/\mdm)^2}{E}\\
&\times\left\{-1 + \ln\left(4(1-E/\mdm)\right) -2\ln\left(\frac{m_l}{\mdm}\right)\right\}\, ,
\end{split}
\end{equation}
in which $\alpha_{\mbox{{\footnotesize EM}}}$ is the electromagnetic
coupling constant and $m_l$ is the mass of the leptons created in the
process $\chi\chi \rightarrow l^+l^-\gamma$. Integrating this spectrum
using an energy range appropriate for \Fermi\ we find that
\begin{equation} 
\int_{4 \mbox{ {\footnotesize GeV}}}^{250 \mbox{ {\footnotesize GeV}}}\dd E\, \frac{\dd
\Ngamma}{\dd E} \approx 0.5\, .
\end{equation}
We find a similar number if we integrate the spectrum from one
cascade step. It is interesting to note that these photon spectra have
fewer photons in them, about an order of magnitude less, than the
annihilation spectra from neutralino annihilation which are often
used \cite{1998APh.....9..137B,2004PhRvD..70j3529F}. Of course, since
the energy spectrum contributes a multiplicative factor to the \DM\
annihilation flux, the effect of different annihilation spectra can be
incorporated by simply multiplying our results with the appropriate
factor.

\subsection{The Dark Matter profile}\label{sec:DMprofile}

Since the photon flux from \DM\ annihilations depends on the density
squared of the \DM\, the exact form of the inner density profile has
important ramifications for the detectability of \DM\
annihilation. The density profile of dark matter halos has been
discussed extensively in the last few years. While the shape of the
outer density profile seems to have converged in the various numerical
simulations of \DM\ halos and a consensus has been reached, the
detailed shape of the inner density profile is still a matter of
debate. The main point of contention is whether the density profile is
cusped near the center of the \DM\ halo, and if so, how steep this
cusp is, or whether the density asymptotes to a constant value at the
center. Since this discussion has a non-negligible impact on the
prospects for \DM\ annihilation detection, we will briefly discuss the
various plausible density profiles.

The inner structure of \DM\ halos is one of the key predictions of the
\CDM\ paradigm, and over a decade ago it was shown that the
spherically averaged density profiles of \DM\ halos are well described
by a universal profile, the so-called \NFW\ profile
\cite{1995MNRAS.275..720N,1996ApJ...462..563N,1997ApJ...490..493N},
depending only on the virial mass of the halo and its
concentration. Moreover, the concentration was shown to depend
systematically on the halo mass
\cite{1996ApJ...462..563N,1997ApJ...490..493N,2001MNRAS.321..559B,2001ApJ...554..114E,2007MNRAS.381.1450N,2008MNRAS.387..536G}. This
profile was found to provide a good fit to many numerical simulations
(\eg,
\cite{1996MNRAS.281..716C,1997MNRAS.286..865T,1997ApJS..111...73K}).
While there seems to be a consensus between different numerical
simulations on the shape of the outer density profile, no agreement
was reached on the inner density structure of the \DM\ halos found in
the simulations. The \NFW\ profile has an inner slope of -1, other
studies, however, found evidence for steeper inner slopes
\cite{1997ApJ...477L...9F,1998ApJ...499L...5M,1999MNRAS.310.1147M,2000ApJ...544..616G,2001ApJ...554..903K,2001ApJ...557..533F,2004MNRAS.353..624D},
shallower inner slopes \cite{1998ApJ...502...48K}, or didn't find any
convergence toward a power-law behavior
\cite{2003MNRAS.338...14P}. Even the universality of the density
profile has been brought into question
\cite{2000ApJ...529L..69J,2000ApJ...535...30J}.

More recently, high-resolution numerical simulations have essentially
ruled out the steepest cusps, such as Moore's profile, which has an
inner cusp a with logarithmic slope of -1.5, however steeper inner
cusps than the original \NFW\ profile are still observed in numerical
simulations \cite{2005MNRAS.364..665D}. Other recent simulations find
inner slopes that are shallower than an \NFW\ profile
\cite{2008MNRAS.385..545K,2008arXiv0808.2981S,2008arXiv0810.1522N}. It
also has become apparent in recent years that the inner density
profile of \DM\ halos might not be described by a power-law
divergence, but that, on the contrary, the logarithmic slope becomes
ever shallower, leading to a cored density profile
\cite{2004MNRAS.349.1039N,2005ApJ...624L..85M,2006AJ....132.2685M,2006AJ....132.2701G,2006MNRAS.365..147S,2008arXiv0808.2981S,2008MNRAS.391.1685S,2008arXiv0810.1522N}. Therefore,
density profiles such as the Einasto profile
\cite{einasto65a,1989A&A...223...89E} have been shown to provide an
excellent fit to the numerical simulations with the highest resolution
yet \cite{2008MNRAS.391.1685S,2008arXiv0808.2981S}.

The implications of the exact form of the inner density profile have
been widely discussed before in the context of various \DM\
annihilation models (\eg,
\cite{1998APh.....9..137B,1999PhRvL..83.1719G,2000PhRvD..62l3005C,2000PhLB..494..181G,2002PhRvD..66b3509T,2002PhRvD..66l3502U,2003MNRAS.339..505T,2004ApJ...601...47A,2008JCAP...07..013B,2008arXiv0811.3744B,2008arXiv0812.3895B}). Roughly
speaking, the annihilation flux is smaller for inner density profiles
which are cored, and larger for inner density profiles which show a
cuspy behavior. The steeper the slope of the inner cusp, the larger
the \DM\ annihilation flux (for obvious reasons). Since the core
vs.~cusp debate is still undecided we will consider \DM\ distribution
models of both types.

The cuspy \DM\ profile we will use is a generalized \NFW\ profile (\GNFW):
\begin{equation}\label{eq:NFW}
\rho(r) = \frac{\rhos}{\redr^\gamma\left(1+\redr\right)^{3-\gamma}};\qquad \redr = r/\rs\, ,
\end{equation}
in which \rhos\ is a characteristic density and \rs\ a characteristic
(scale-)radius.  The original \NFW\ profile is recovered for $\gamma$
= 1 \cite{1997ApJ...490..493N}, which we will use as the fiducial
model in what follows. Varying the value of the inner logarithmic
slope $\gamma$ will allow us to describe the effect of steeper and
shallower inner slopes than the \NFW\ model. Note that this functional
form does \emph{not} include the exact form of Moore's profile; this
need not concern us since inner cusps as steep as Moore's profile are
excluded by the latest numerical simulations.

\begin{figure}[t]
\centering
\includegraphics{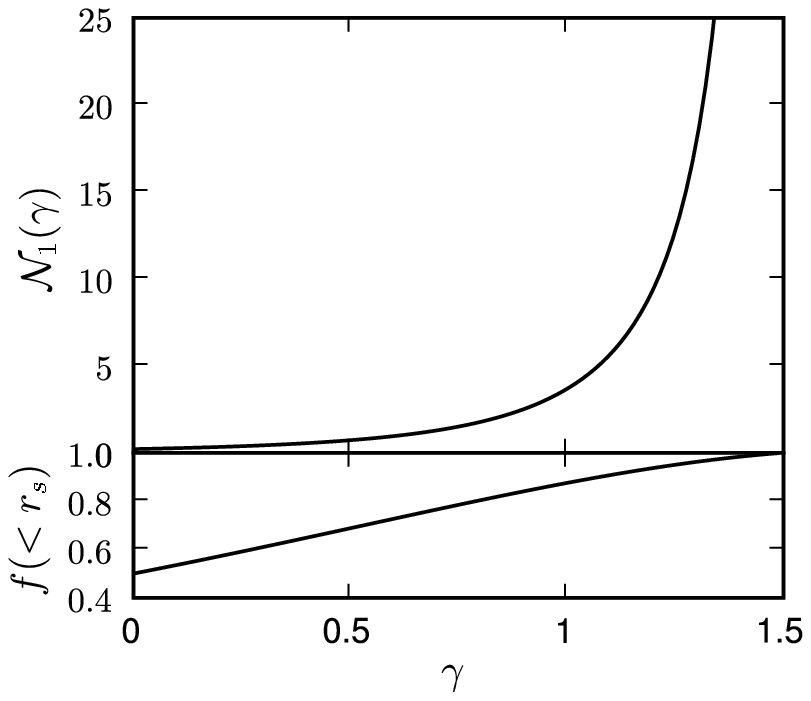}
\caption{Top: Normalization of the $\lum \propto \rhos^2\,
\rs^3/\dist^2$ relation of \eqnname\ (\ref{eq:lumrhosrsNFW}) for the
\GNFW\ profile. The functional dependence of the normalization on the
exponent $\gamma$ in the profile given in \eqnname\ (\ref{eq:NFW}) is
given in \eqnname\ (\ref{eq:normgammaNFW}). Bottom: Fraction of the
emission that comes from the region within \rs.}%\
\label{fig:normgammaNFW}%
\end{figure}

The contribution of the \DM\ density profile on the structure quantity
\lum\ can be simply evaluated in the context of any density model. As
in Ref.~\cite{2007PhRvD..75h3526S} we assume that the distance to the
subhalo \dist\ is much larger than the scale-radius of the \DM\
halo. For an \NFW\ profile 90\% of the flux originates from the region
within one scale-radius, and we will, conservatively, only take
contributions to the \DM\ annihilation flux coming from this region
for all of the different profiles. Therefore, we find that
\begin{equation}\label{eq:lumrhosrsNFW}
\int_{\Delta\Omega}\,\frac{\dd\Omega}{4\pi}\,\int_{\mathrm{\los}} \dd
l\, \rho^2(l)= \normnfw(\gamma)\, \frac{\rhos^2\, \rs^3}{\dist^2}\, .
\end{equation}
The normalization of this relationship depends on the logarithmic
slope of the inner cusp and is given by
\begin{equation}\label{eq:normgammaNFW}
\normnfw(x) = 4 \pi  \, \frac{2^{2x-5}\left(-16+11x-2x^2\right)}{-30+47x-24x^2+4x^3}\, .
\end{equation}
The normalization for the \NFW\ profile is $7\pi/6$ and it depends
strongly on the value of the logarithmic slope of the inner density
profile, as can be seen in Figure \ref{fig:normgammaNFW}. The bottom
panel of Fig.~\ref{fig:normgammaNFW} shows the fraction of the \DM\
annihilation flux coming from region within one scale-radius of the
halo. For reasonable values of the inner slope this fraction is in the
range 80 to 95\%.

\begin{figure}[t]
\centering
\includegraphics{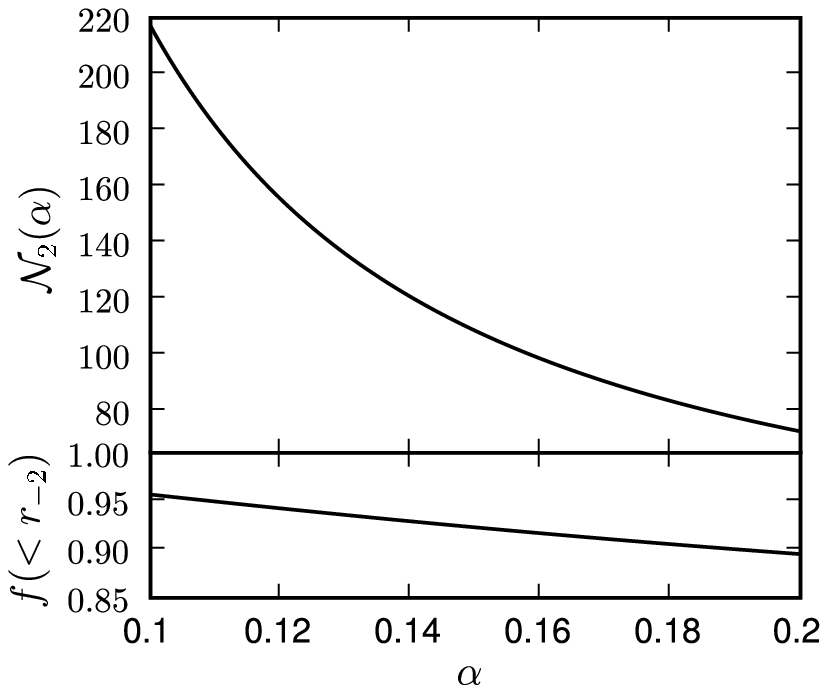}
\caption{Top: Normalization of the $\lum \propto \rhominustwo^2\,
\rminustwo^3/\dist^2$ relation of \eqnname\
(\ref{eq:lumrhosrsEinasto}) for the Einasto profile. The functional
dependence of the normalization on $\alpha$ in the profile given in
\eqnname\ (\ref{eq:einasto}) is given in \eqnname\
(\ref{eq:normgammaEinasto}). Bottom: Fraction of the emission that
comes from the region within \rminustwo.}%\
\label{fig:normgammaEinasto}%
\end{figure}

The Einasto profile is characterized by an ever-decreasing logarithmic
inner slope. It's functional dependence on the radius is given by
\begin{equation}\label{eq:einasto}
\rho (r) = \rhominustwo\, \exp\left[-\frac{2}{\alpha}\left\{\left(\frac{r}{\rminustwo}\right)^{\alpha}-1\right\}\right]\, .
\end{equation}
Here, \rhominustwo\ is the density at the radius \rminustwo\ where the
local slope is -2 and \alphaEinasto\ is a shape parameter often fixed
at a value $\sim\!0.17$ \cite{2008MNRAS.391.1685S}. As above, assuming
$\dist \gg \rminustwo$, calculating the flux originating from within
the radius \rminustwo, we find that
\begin{equation}\label{eq:lumrhosrsEinasto}
\int_{\Delta\Omega}\,\frac{\dd\Omega}{4\pi}\,\int_{\mathrm{\los}} \dd
l\, \rho^2(l)= \normeinasto(\alpha)\, \frac{\rhominustwo^2\, \rminustwo^3}{\dist^2}\, .
\end{equation}
The normalization depends on the value of \alphaEinasto\ and is given
by
\begin{equation}\label{eq:normgammaEinasto}
\normeinasto(x) = 4 \pi \, 2^{-6/\alpha}\,e^{4/\alpha}\,
\alpha^{-1+3/\alpha}\,
\gamma\left[\frac{3}{\alpha},\frac{4}{\alpha}\right]\,
,
\end{equation}
where $\gamma[a,z]$ is the lower incomplete gamma function. The
dependence of this normalization on \alphaEinasto\ is shown in Figure
\ref{fig:normgammaEinasto}. In order to compare the Einasto profile as
described here to the \GNFW\ profile, we can relate the parameters
$(\rhos,\rs)$ of the \GNFW\ profile to the parameters
$(\rhominustwo,\rminustwo)$ of the Einasto model by calculating the
radius at which the logarithmic slope of the density versus radius
profile equals -2. Thus, we find that
\begin{equation}
\rminustwo = \rs,  \qquad \qquad \rhominustwo = \frac{\rhos}{2^{3-\gamma}}\, .
\end{equation}

Note that the \DM\ annihilation flux from the Einasto profile,
although it seems larger from Figs.~\ref{fig:normgammaNFW} and
\ref{fig:normgammaEinasto}, is actually smaller in general, since
there is a factor of 2$^{6-2\gamma}$ involved in going from one to the
other.

Further on it will sometimes be useful to describe the \DM\ density
profiles by different parameters than a characteristic density and
radius. Another set of parameters is given by a concentration
parameter and a mass. The virial radius \Rvir\ of a halo is defined to
be the radius within which the mean density of the halo is $\Delta$
times the critical density $\rhoc$, where $\Delta = 178\,
\omegam^{0.45} \approx 100$
\cite{1996MNRAS.282..263E,1998ApJ...495...80B}. The virial mass is
simply the mass contained within a sphere of radius \Rvir
\begin{equation}\label{eq:virialmass}
\Mvir \equiv \frac{4\pi}{3} \Delta \rhoc \Rvir^3\, .
\end{equation}
The concentration for both the \GNFW\ profile as for the Einasto
profile is then defined by
\begin{equation}\label{eq:concdef}
\conc \equiv \frac{\Rvir}{\rminustwo}\, .
\end{equation}
A characteristic mass of a halo is given by the virial mass defined
above. The virial mass of the \GNFW\ profile in terms of the
characteristic density \rhos\ and the characteristic radius \rs\ is
\begin{equation}\label{eq:Mvirdef}
\Mvir = 4\pi\rhos\rs^3 f(c)\, ,
\end{equation}
in which the function $f(c)$ is given by
\begin{equation}\label{eq:hypergeometric}
f(c) = \frac{c^{3-\gamma}}{3-\gamma} \, {_2F_1}\left(3-\gamma,3-\gamma;4-\gamma;-c\right)\, .
\end{equation}
The function $_2 F_1\left(a,b;c;z\right)$ is the Gauss hypergeometric
function. In the $\gamma = 1$ case $f(c)$ is given by the familiar
form $f(c) = \ln(1+c)-c/(1+c)$. The virial mass for the Einasto
profile in terms of the characteristic density \rhominustwo\ and the
characteristic radius \rminustwo\ is
\begin{equation}
\Mvir = 4\pi\rhominustwo\rminustwo^3 \frac{1}{\alphaEinasto} \exp\left(\frac{3\ln \alphaEinasto + 2 - \ln 8}{\alphaEinasto}\right) \gamma\left[\frac{3}{\alphaEinasto},\frac{2}{\alphaEinasto}\conc^\alphaEinasto\right]\, .
\end{equation}
In this last expression $\gamma\left[a,z\right]$ is the lower
incomplete gamma function. The density can then be expressed in terms
of the concentration by using the definition of the virial mass given
in \eqnname~(\ref{eq:virialmass}).

Another pair of parameters that one can use to characterize the \DM\
density profile is the maximum circular velocity \vmax\ and the radius
at which this maximum occurs, \rmax. These parameters are especially
well-suited when dealing with numerical simulations, since they can be
readily read off from the raw simulation data. The circular velocity
is given by
\begin{equation}
\vcirc^2(r) \equiv \frac{G M(r)}{r} = \vcirc^2(\Rvir)\frac{c}{f(c)}\frac{f(x)}{x}\, ,
\end{equation}
in which $x \equiv r/\rminustwo$ and $f(x)$ is defined in \eqnname\
(\ref{eq:hypergeometric}) in the case of a \GNFW\ profile and is given
by
$\gamma\left[\frac{3}{\alphaEinasto},\frac{2}{\alphaEinasto}x^\alphaEinasto\right]$
in the case of an Einasto profile. One can find \rmax\ by
(numerically) solving the equation
\begin{equation}
\frac{\dd}{\dd x}\left[ \frac{f(x)}{x}\right] = 0\, ,
\end{equation}
and \vmax\ follows then immediately.

\subsection{Calculating the total boost}

As is customary, we will write the total structure quantity \lum($M$) as a
sum of a smooth component and a substructure component:
\begin{equation}
\lum(M) = \lumsmooth(M) + \lumsub(M)\, .
\end{equation}
The smooth contribution to the structure quantity is the product of
the Sommerfeld enhancement \somm\ and the line-of-sight integral over
the \DM\ density squared, given for the different \DM\ density
profiles in \eqnname s (\ref{eq:lumrhosrsNFW}) and
(\ref{eq:lumrhosrsEinasto}). Therefore,
\begin{equation}
\lumsmooth(M) = \somm \, \lumsmoothc(M)\, ,
\end{equation}
in which \lumsmoothc\ is the luminosity that would come from the
smooth component of the halo in the absence of Sommerfeld enhancement.

We can now write
\begin{equation}\label{eq:totboost}
\begin{split}
\lum(M) & = \somm\,\lumsmoothc(M) + \lumsub(M)\\
&= \left[ \somm + \frac{\lumsub(M)}{\lumsmoothc(M)}\right] \lumsmoothc(M)\, .
\end{split}
\end{equation}
We will from now on write the substructure boost as \boost(M). Then we
have the following
\begin{equation}\label{eq:boost}
\begin{split}
\boost(M) & \equiv \frac{\lumsub(M)}{\lumsmoothc(M)} = \frac{1}{\lumsmoothc(M)} \int \dd m \, \frac{\dd N}{\dd m} \lum(m)\\
&= \frac{1}{\lumsmoothc} \int \dd m\, \frac{\dd N}{\dd m} \left[ \somm(m) + \boost(m) \right] \lumsmoothc(m)\, .
\end{split}
\end{equation}
Here we have introduced the subhalo mass function $\dd N/\dd m$, which
gives the abundance of subhalos of mass $m$.

Before proceeding we should specify the upper and lower limits of the
integral appearing in \eqnname\ (\ref{eq:boost}). The integration over
ever smaller substructures is cut-off at the size of the smallest
subhalos. The mass scale at which this cut-off occurs is the thermal
free-streaming scale, a \DM\ model-dependent quantity, which in many
\DM\ models lies in the range 10$^{-6}$ to 10$^{-12}$ \Msol\
\cite{1999PhRvD..59d3517S,2001PhRvD..64h3507H,2004MNRAS.353L..23G,2005PhRvD..71j3520L,2006PhRvL..97c1301P}. The
upper limit to this integral is some fraction $q$ of the halo mass
$M$, since the abundance of substructure does not continue up to this
mass scale. In the following we will set $q$ equal to 0.1.

In order to calculate the total boost from substructure to the
annihilation cross section, we will write all of the quantities
appearing in \eqnname\ (\ref{eq:boost}) as a function of the mass
$m$. For instance, from numerical simulations, the subhalo mass
function is well constrained to follow a power-law type behavior:
\begin{equation}\label{eq:massfraction}
\frac{\dd N}{\dd \ln m}
\propto \left(\frac{m}{M}\right)^n\, .
\end{equation}
The value of the exponent $n$ is still under debate. The two highest
resolution $N$-body simulations yet, the \aquarius\ and \VL\
projects, both find values for $n$ that put roughly equal amounts of
mass per subhalo mass decade. While the \VL\ simulation seems to
prefer a value $n = -1$
\cite{2006ApJ...649....1D,2008ApJ...686..262K}, which would lead to a
logarithmically diverging total mass in substructures if the subhalo
abundance were not cut off at the thermal free-streaming limit of the
dark matter, the \aquarius\ project prefers a value $n = -0.9$
\cite{2008MNRAS.391.1685S}. As the \aquarius\ project has made
detailed convergence studies that support their claim for a value of
$n$ that is shallower than -1, we will adopt the value $n = -0.9$ as
our fiducial value, although even shallower slopes have also been
observed \cite{2002PhRvD..66f3502H}. The same slope has also been
observed in other simulations \cite{2004MNRAS.355..819G}.

With the exponent $n$ fixed, we turn to the normalization of the
relation given in \eqnname\ (\ref{eq:massfraction}). Again, different
high-resolution numerical simulations give different answers to this
question as it relates to the abundance of subhalos of Galactic
subhalos. The problem of the normalization is complicated by the fact
that numerical simulations have only recently begun to be able to
resolve substructure in subhalos in a Galaxy-size halo. While the
simulations generally agree that the resolved substructure abundance
of subhalos in the host halo adds up to $\sim\!10$\% of the total mass
of the halo
\cite{1998MNRAS.300..146G,1999ApJ...522...82K,2004MNRAS.353..624D,2003ApJ...598...49Z,2007ApJ...657..262D,2007ApJ...667..859D},
the question of whether the substructure abundance in subhalos is just
a scaled-down version of the substructure abundance in the host halo
remains unresolved. One would not expect the abundance to be
self-similar, since, while substructure in both the main host halo and
subhalos gets diminished by tidal disruptions, substructure in the
subhalos does not get replenished by the infall of halos from the
field, as is the case for the host halo. Nevertheless, many numerical
simulations do find approximately self-similar substructure abundances
\cite{1999ApJ...524L..19M,2007ApJ...659.1082S}. Recently, the
\aquarius\ simulation has concluded from its simulation that the
substructure abundance is not self-similar, although the deviation
from the self-similar relation is small
\cite{2008MNRAS.391.1685S}. Therefore, we will adopt a normalization
here which places 10\% of the mass of the subhalo in the form of
``resolved'' substructure, where we adopt the definition that resolved
substructure is in halos of mass $\gtrsim 10^{-5} M$.

In the previous section we calculated $\lumsmoothc$ for the different
\DM\ density profiles and found in all cases that
\begin{equation}
\lumsmoothc(m) \equiv \lumsmoothc(\rho(m),r(m))\, ,
\end{equation}
for a characteristic density $\rho$ and radius $r$, the exact
definition of which depends on the specific \DM\ density profile
used. In order to find the mass dependence of $\lumsmoothc$, we use
the description of the \DM\ density profile in terms of virial mass
\Mvir\ and concentration \conc. We then make use of the fact that many
simulations have found a relation between the concentration of halos
and their mass
\cite{2001MNRAS.321..559B,2001ApJ...554..114E,2008MNRAS.387..536G,2007MNRAS.381.1450N},
such that everything depends only on the mass. This relation is well
established over a large range in halo masses, and we will adopt the
particular relationship found in Ref.~\cite{2007MNRAS.378...55M}
\begin{equation}\label{eq:concscaling}
\conc = 10.5 \, \left(\frac{\Mvir}{10^{12}\, \Msol}\right)^{-0.11}\, ,
\end{equation}
in good agreement with the results of other, similar analyses
\cite{2001MNRAS.321..559B,2005MNRAS.357..387K}, although the
normalization of the concentration-mass relation is about 15\% lower
than the original relation. However, the concentration of subhalos is
generally larger than the concentration of field halos of the same
mass as a result of tidal mass-loss
\cite{1998MNRAS.300..146G,2008ApJ...673..226P,2004MNRAS.355..794H,2005ApJ...635..931B,2004ApJ...608..663K}. This
is a small effect, of order 10\%, and it does not play a large role in
the results we obtain below. This concentration-halo mass
concentration has not just been established in numerical simulations:
statistical weak lensing analysis of luminous red galaxies and
clusters from the Sloan Digital Sky Survey (\SDSS) has shown that the
halo profiles of these objects are consistent with the \NFW\ profile
and mass-concentration relation given above
\cite{2008JCAP...08..006M}.

The only remaining quantity in \eqnname\ (\ref{eq:boost}) that we
haven't related to the mass of the subhalo yet, is the Sommerfeld
enhancement factor \somm. The Sommerfeld enhancement depends on the
mass of the subhalo through the velocity dispersion. Therefore, we
need to relate the velocity dispersion of a \DM\ subhalo to its
mass. It is important to note that all we really need is an
approximate relationship between the velocity dispersion and the halo
mass which fits well the range of Galactic size halos to a scale not
much smaller than the mass scale of the smallest dwarf Spheroidals
known today, since in most cases the Sommerfeld enhancement will
saturate at velocity dispersions of the order of the velocity
dispersion of the dwarf Spheroidals. This means that the specific form
that we adopt below for the velocity dispersion-mass scaling relation
does not matter greatly.

\begin{figure*}[t]
\includegraphics[clip=]{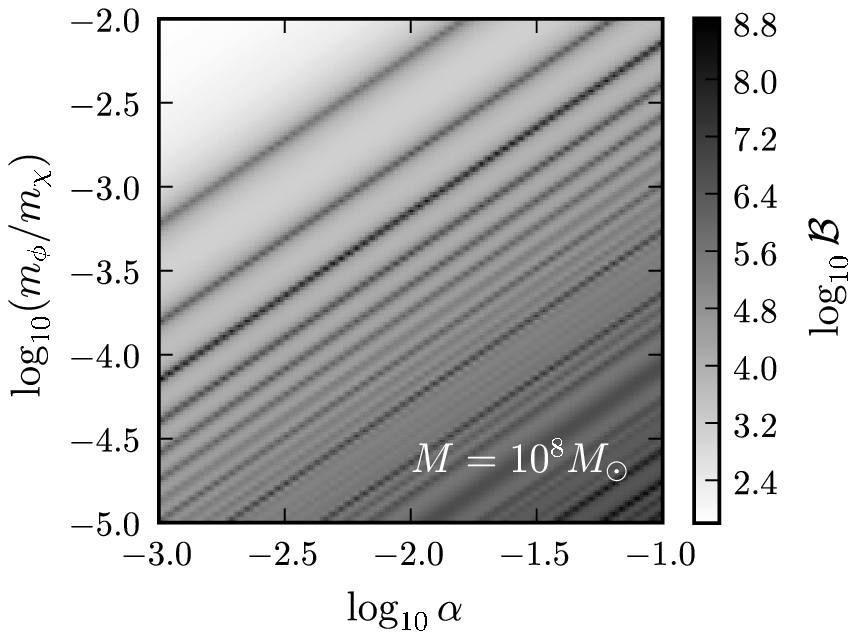}%%BoundingBox: 178 291 400 474
\includegraphics[clip=]{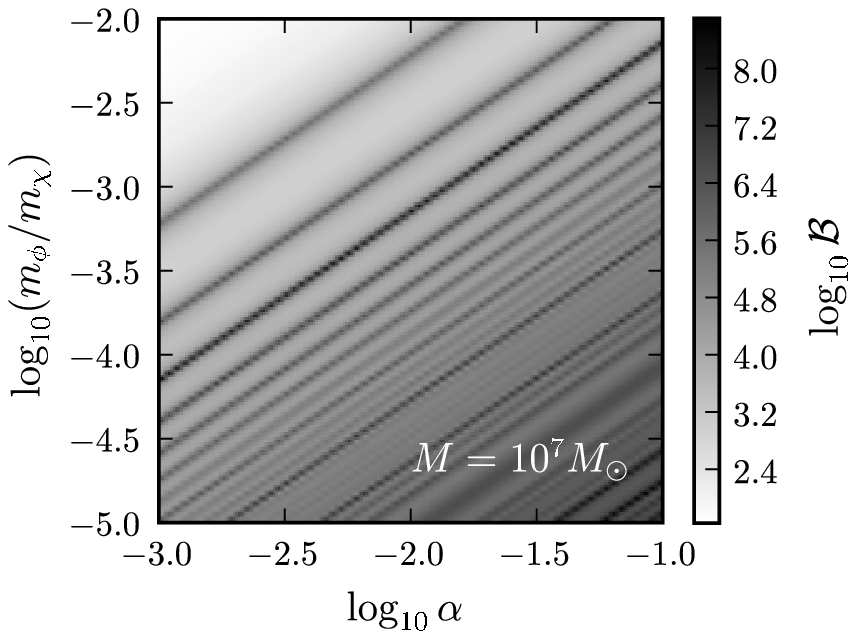}\\%%BoundingBox: 190 291 430 474 
\includegraphics[clip=]{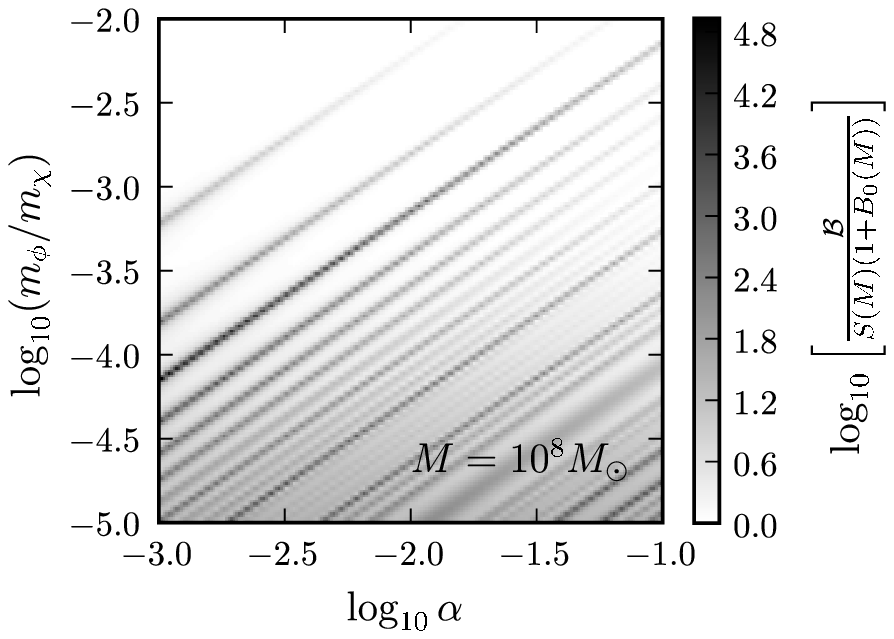}%%BoundingBox: 170 277 398 487 
\includegraphics[clip=]{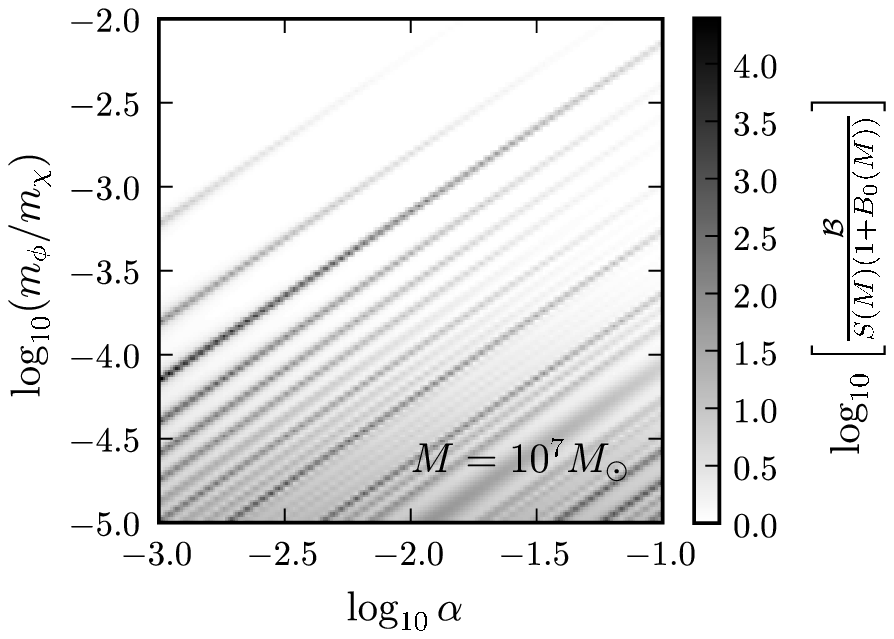}%%BoundingBox: 192 277 432 487
\caption{The total (defined in Eq.~[\ref{eq:totalboost}]) and reduced
boost (defined in Eq.~[\ref{eq:redboost}]) for subhalos of two
different mass scales. The dark matter profile of all the subhalos is
assumed to be well-described by an \NFW\ profile.}%
\label{fig:subboost}%
\end{figure*}

\begin{figure*}[t]
\includegraphics[clip=]{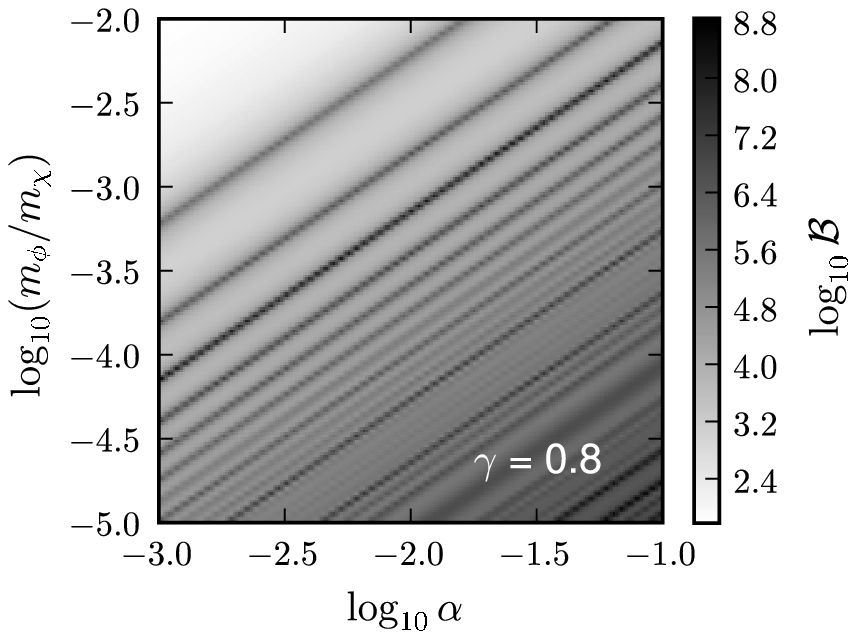}%%BoundingBox: 178 291 400 474
\includegraphics[clip=]{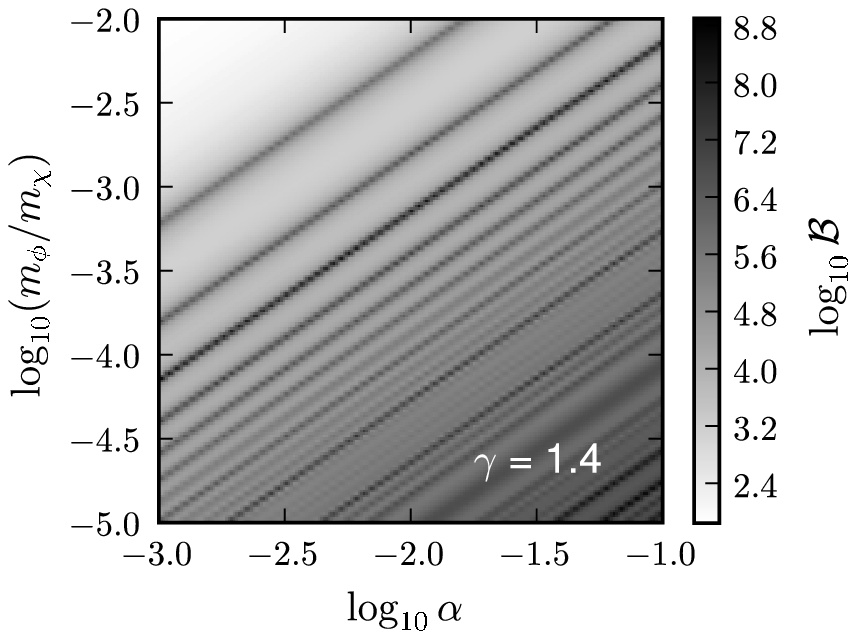}\\%%BoundingBox: 190 291 430 474
\includegraphics[clip=]{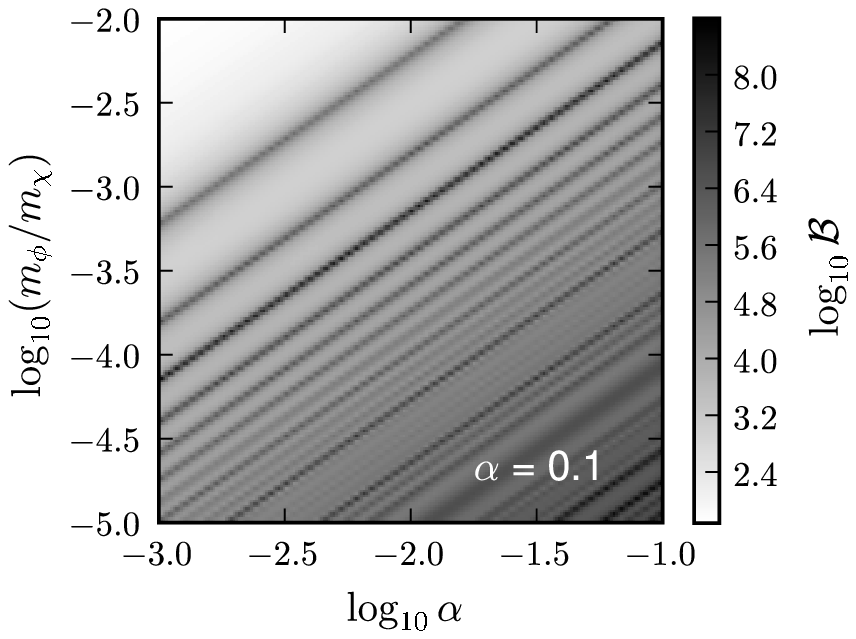}%%BoundingBox: 178 291 400 474
\includegraphics[clip=]{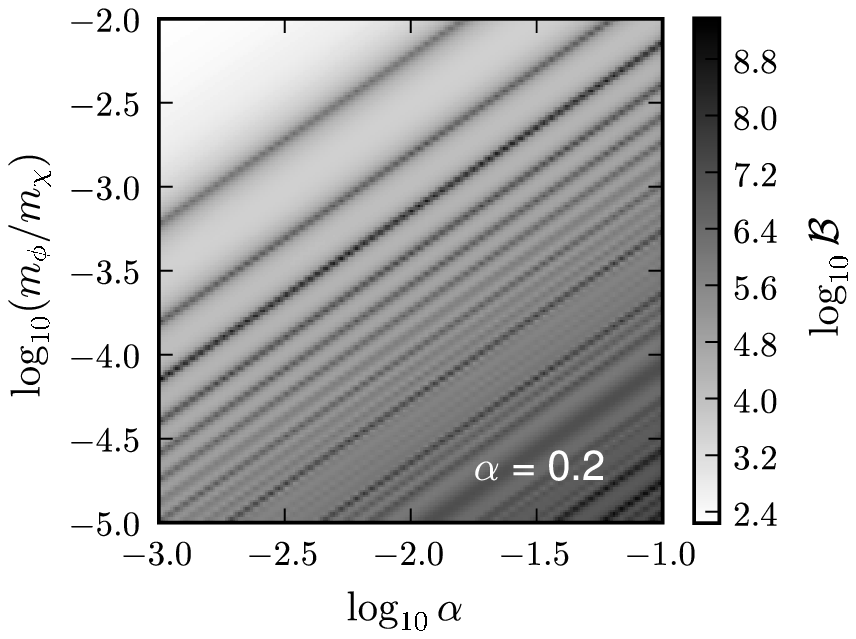}%%BoundingBox: 190 291 430 474
\caption{The total boost for a $M=10^8$ \Msun\ subhalo for different
shapes of the \DM\ density profile: a generalized \NFW\ profile
described by an inner slope $\gamma$ (\emph{top panels}); an Einasto
profile described by a shape parameter \alphaEinasto\ (\emph{bottom
panels}).}%
\label{fig:subboostprofiles}%
\end{figure*}

\begin{figure*}[t]
\includegraphics[clip=]{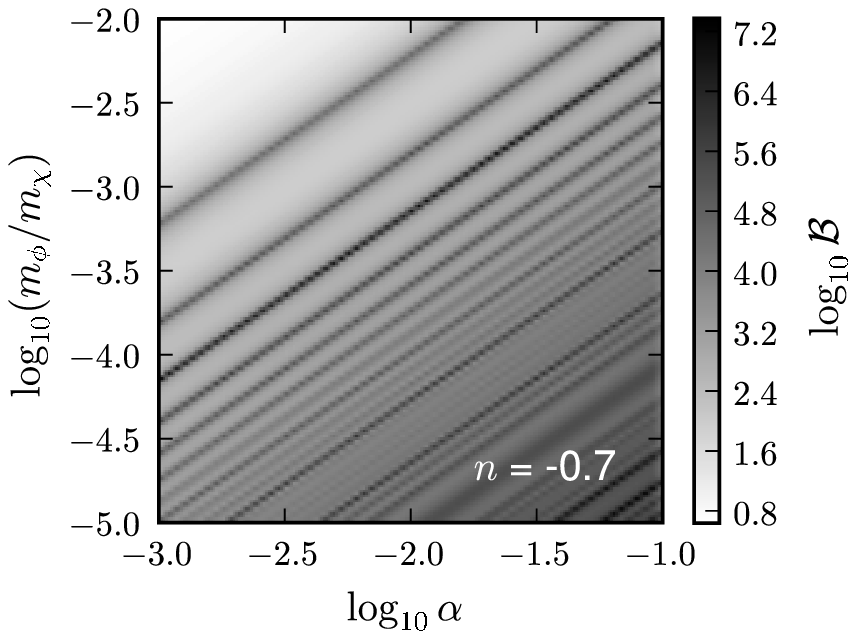}%%BoundingBox: 178 291 400 474
\includegraphics[clip=]{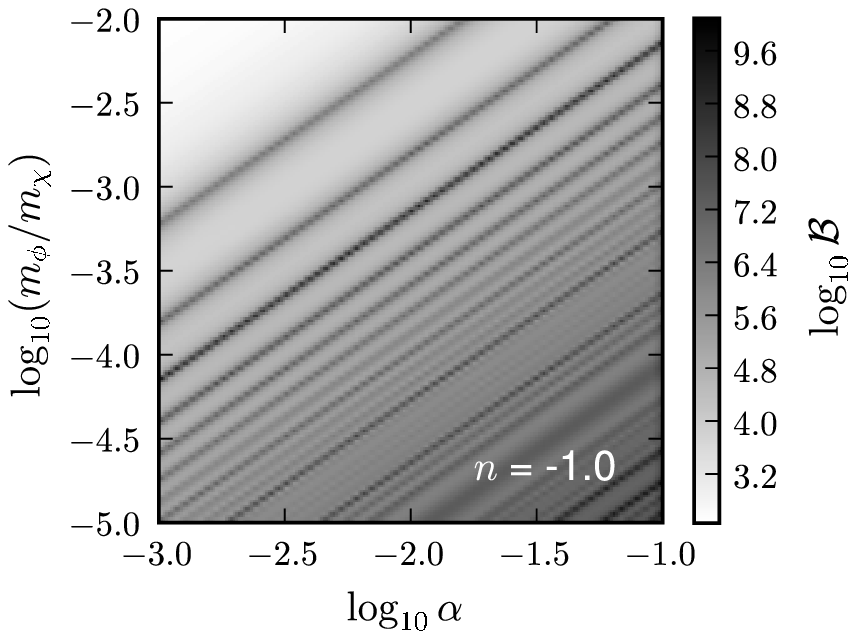}\\%%BoundingBox: 190 291 430 474
\includegraphics[clip=]{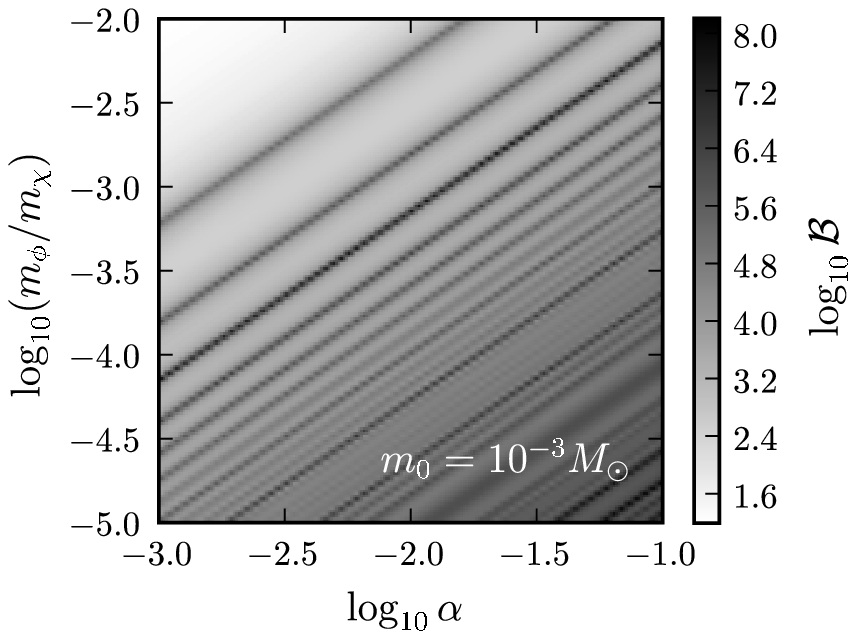}%%BoundingBox: 178 291 400 474
\includegraphics[clip=]{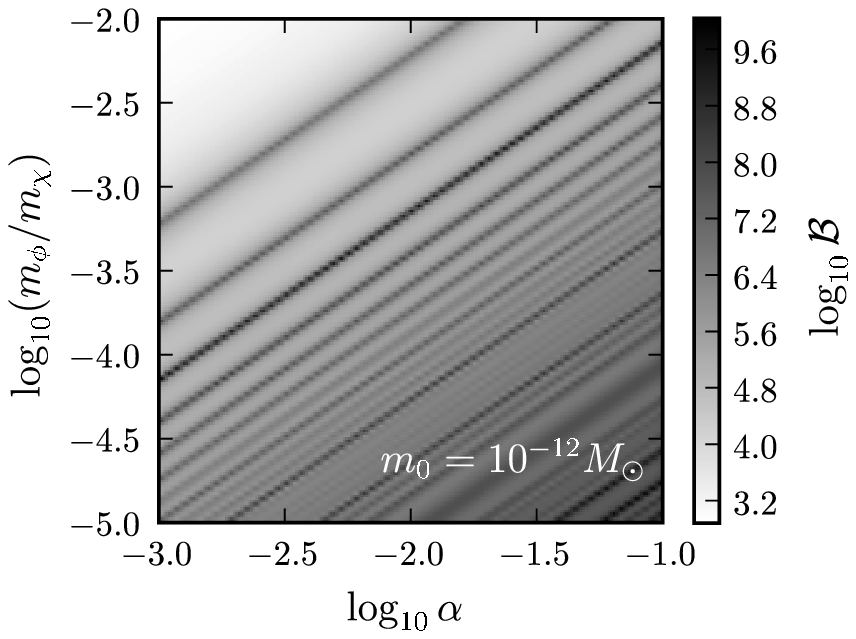}%%BoundingBox: 190 291 430 474
\caption{The total boost for a $M=10^8$ \Msun\ for different slopes
$n$ for the subhalo mass function (\emph{top panels}) and for
different values of the dark matter thermal free streaming limit
$m_0$ (\emph{bottom panels}).}%
\label{fig:subboostslope}%
\end{figure*}

On the scale of clusters there is a well-defined relationship between
the mass of a \DM\ halo and its one-dimensional velocity dispersion
which one can derive from adiabatic scaling arguments
\cite{1998ApJ...495...80B} given by
\begin{equation}\label{eq:vdispscaling}
\sigma \propto m^{1/3}\, .
\end{equation}
This scaling relation can be easily derived in the case of an
isothermal distribution function \cite{2008gady.book.....B}, since
then
\begin{equation}
\rho(r) = \frac{\sigma^2}{2\pi G r^2}\, ,
\end{equation}
from which \eqnname\ (\ref{eq:vdispscaling}) immediately follows. This
scaling relation has been found to describe numerical simulations of
cluster-sized halos extremely well
\cite{1991ApJ...383...95E,1995MNRAS.275..720N,1995AJ....110...21C,1996ApJ...469..494E,1996MNRAS.281..716C,1998ApJ...495...80B}. From
recent numerical simulations it seems that we can use this relation
even for subhalos of subhalos in a Galaxy sized dark matter halo. In
Ref.~\cite{2008MNRAS.391.1685S} it is shown that the relation between
the maximum circular velocity of a subhalo and its mass is well fit by
a power-law of the type
\begin{equation}
m \propto \vmax^{3.5}\, .
\end{equation}
Using the circular velocity at \rminustwo\ as a measure of the halo's
velocity dispersion gives $\sigv \propto m^{1/3.5}$. Alternatively, we
could use the circular velocity at the virial radius as a measure of
the velocity dispersion. In the case of an \NFW\ profile we have that
\begin{equation}
\frac{\vmax^2}{\vcirc(\Rvir)^2} \propto \frac{c}{f(c)}\, ,
\end{equation}
in which $f(x) = \ln(1+c) - c/(1+c) \approx 2.6 (c/33)^{0.4}$ in the
relevant concentration range. Using \eqnname\ (\ref{eq:concscaling}),
this gives
\begin{equation}
\sigv \propto m^{0.32}\, .
\end{equation}
Therefore, we will use (as a rough approximation) the scaling relation
given in \eqnname\ (\ref{eq:vdispscaling}). We calibrate this
relationship at the mass scale of the dwarf spheroidals by using the
velocity and mass of Canes Venatici I from Ref.~\cite{2007ApJ...670..313S}.

\subsection{The total and the ``reduced'' boost}

We define the total boost for a given subhalo mass as (see \eqnname\
[\ref{eq:totboost}])
\begin{equation}\label{eq:totalboost}
\totalboost(M) = \somm(M) + \boost(M)\, ,
\end{equation}
in which $\somm(M)$ is the Sommerfeld enhancement for a halo of mass
$M$ and $\boost(M)$ is the substructure boost for a halo of that mass,
as defined in Eq.~(\ref{eq:boost}). This total boost for subhalos of
mass $M$ = $10^8$ \Msun\ and $M$ = $10^7$ \Msun\ calculated by
following the procedure outlined in the previous subsection is shown
in Figure \ref{fig:subboost} (\emph{top panels}). It is immediately
obvious that the total boost factors can get very large in certain
regions of the parameter space spanned by \mv/\mdm and $\alpha$:
boosts vary between $10^2$ to $10^9$. These masses correspond to the
typical masses of the classical and newly-discovered dwarf Spheroidals
(\dSphs) and as such their velocity dispersions are of the order
$10^{-5}$. The pattern of resonance regions and non-resonance regions
that appears for the pure Sommerfeld enhancement (as shown in Figure
\ref{fig:sommavg}) is also apparent in these plots of the total
boosts, however, it is noteworthy that the resonances seems to be more
highly peaked than they were in the pure Sommerfeld enhancement case,
and that the boost the lower half of the parameter space, \ie, for
smaller values of \mv/\mdm, are higher than one would expect from the
pure Sommerfeld enhancement combined with a boost factor from pure
density enhancements.

This is confirmed by looking at the ``reduced boost'' defined as
\begin{equation}\label{eq:redboost}
\redboost(M) \equiv \frac{\totalboost(M)}{\somm(M)\left(1+\boost_0(M)\right)}\, ,
\end{equation}
in which $\boost_0(M)$ is the boost we would find if we turned off the
Sommerfeld enhancement in the calculation of the total boost, \ie, the
boost we would get if we set $\somm(m)$ equal to one in \eqnname\
(\ref{eq:boost}). This $\boost_0$ corresponds to the boost factor that
has been used before to estimate the total boost from density
enhancements in \dSphs\ and it is generally \order(10). This reduced
boost therefore divides out the contribution of the Sommerfeld
enhancement of the top-level subhalo and the boost from the
overdensities in lower-level halos, and it would be equal to one if
there were no extra boost coming from the fact that the substructures
are kinematically colder than the main subhalo. As we can see in
Figure \ref{fig:subboost} (\emph{bottom panels}), in various parts of
the Yukawa coupling parameter space this reduced boost is equal to one
such that there is no extra boosts caused by the lower velocity
dispersion in the substructures. This is mainly the case for large
values of the parameter \mv/\mdm, since the Sommerfeld enhancement, in
the 1/$v$ regime, levels off around $\somm \sim \mv/\mdm$, and since
the top-level subhalo has a velocity dispersion of $\sim 10^{-5}$ this
means that the Sommerfeld enhancement has mostly saturated at these
large values of \mv/\mdm, even in regions of parameter space in which
$\alpha$ is small. Indeed, in the 1/$v$ regime the Sommerfeld
enhancement is $\sim \alpha/\beta$ $\approx$ $10^{2}$ for $\alpha =
10^{-3}$, such that the Sommerfeld enhancement has fully saturated
over the whole $\alpha$ range for large values of \mv/\mdm\ ($\mv/\mdm
\sim 10^{-2}$).

However, the Sommerfeld enhancement has not saturated in all regions
of parameter space at this mass scale, and for smaller values of
\mv/\mdm and especially in resonance regions, the reduced boost can
reach high values, signaling a large extra boost caused by the
kinematically coldness of the substructure.  For small values of
\mv/\mdm\ we see that we can get boosts that are an order of magnitude
to two orders of magnitude larger than what we would expect from
naively combining the Sommerfeld enhancement from the velocity
dispersion of the subhalo with the boost from density enhancements
alone. This can be easily understood as being a consequence of the
fact that the Sommerfeld enhancement has not saturated yet at the mass
scale of the top-level subhalo, such that it can be boosted up to it
saturation point by the contribution from the kinematically colder
substructure.

The largest values of the reduced boost are found in the resonance
regions. In these regions the Sommerfeld enhancement both grows faster
with decreasing relative velocity as well as saturates at a much
larger value of the enhancement. Both of these effects conspire to
give very large extra boosts. The resonance regions are clearly
distinguishable in this plot of the reduced boost as 45$\degree$
lines. Because the resonance regions are so narrow the apparent extra
boosts in different resonances seems to vary greatly, but this is
mostly caused by the finite resolution of this figure. Extra boosts up
to $\sim 10^5$ are obtained in these resonances, and these extra
boosts increase along the resonances when decreasing $\alpha$, again
caused by the higher saturation level for smaller $\mv/\mdm$ values.

The total boost factors obtained by assuming an \NFW\ dark matter
density profile do not depend much on the inner slope of the density
profile. This is mainly because the calculation of the boost only
considers \emph{relative} luminosities, for which the difference in
normalization of the $\lum \propto \rhos^2 \rs^3/\dist^2$ relation for
different inner slopes is irrelevant, and because the mass of the
subhalo is only weakly affected by the value of the inner slope---the
function $f(c)$ is only a weak function of the inner slope. This is
confirmed by calculating the total boost for a $M = 10^8$ \Msun\
subhalo for a subhalo with a very shallow inner cusp as well as for a
subhalo with a very steep inner slope. These are shown in Figure
\ref{fig:subboostprofiles} (\emph{top panels}), and, from comparing
this with the total boost shown in Figure \ref{fig:subboost}, it is
clear that the inner slope does not affect the total boost caused by
substructure, neither the overall magnitude of the effect nor the
structure in the (\mv/\mdm,$\alpha$) plane. The main difference in
\DM\ annihilation flux between the different dark matter density
profiles is therefore given by the normalization of the $\lum \propto
\rhos^2 \rs^3/\dist^2$ relation, shown in Figure
\ref{fig:normgammaNFW}.

If the \DM\ density profiles of halos is better described by an
Einasto profile then the substructure boost is affected more strongly
by the shape parameter \alphaEinasto\, which basically describes how
cored the \DM\ density profile is. Changing \alphaEinasto\ has a
greater impact on changing the overall \DM\ density profile and
therefore also plays a larger role in the relation between the
characteristic density, the concentration, and the mass of the
subhalo. The total boosts for a $M$ = $10^8$ \Msun\ subhalo is shown
in Figure \ref{fig:subboostprofiles} (\emph{bottom panels}) for two
different values of the shape parameter \alphaEinasto\, which span the
range found in numerical simulations (for values of \alphaEinasto\
greater than 0.2 the behavior basically saturates at that of
\alphaEinasto\ = 0.2). For \alphaEinasto\ = 0.1 the total boost is
about the same as for the \GNFW\ profile shape, which is
understandable given that the Einasto profile is more cuspy for
smaller values of \alphaEinasto. The same dependence on the parameter
\mv/\mdm\ as is seen for the \GNFW\ profile is displayed for the
Einasto profile. In the case that \alphaEinasto\ is larger the
substructure boost reaches slightly larger values than for the \GNFW\
density profile. Thus, as was the case for the \GNFW\ profile, the
impact of the different normalization of the
$\rhominustwo^2\rminustwo^3/\dist^2$ for different values of
\alphaEinasto\ will play a leading role in the difference between the
different profile shapes.

\begingroup
%\squeezetable
\begin{table*}
\begin{ruledtabular}
\begin{tabular}{lrr@{.}lr@{}lr@{}lr@{.}lr@{.}lr@{.}l@{ $\times$ }lc}
dSph & \dist\ & \multicolumn{2}{c}{$\sigma_0$}\ & \multicolumn{2}{c}{\rmax} & \multicolumn{2}{c}{\vmax} & \multicolumn{2}{c}{\rhos} & \multicolumn{2}{c}{\rs} & \multicolumn{3}{c}{$M(<\rs)$} & References\\
& [kpc] & \multicolumn{2}{c}{[km s$^{-1}$]} & \multicolumn{2}{c}{[kpc]} & \multicolumn{2}{c}{[km s$^{-1}$]} & \multicolumn{2}{c}{[$\Msun$ kpc$^{-3}$]} & \multicolumn{2}{c}{[kpc]} & \multicolumn{3}{c}{[$\Msun$]}&\\[4pt]
\hline
Carina.......................................................................... & 101 & 6&8 & 3&\phantom{f}      & 16&$\!\!\!$      & 1&1 $\times$ $10^7$ & 1&4 & 7&4&$10^8$ & 1,3\\
Coma Berenices...........................................................      & 44  & 4&6 & $\cdot$&& $\cdot$& & 3&0 $\times$ $10^8$ & 0&3 & 2&0&10$^7$ & 2,4\\
Draco........................................................................... & 80  & 5&5 & 8&      & 40&      & 1&0 $\times$ $10^7$ & 3&7 & 1&2&$10^9$ & 1,3\\
Fornax.........................................................................  & 138 & 11&1& 4&      & 20&      & 1&0 $\times$ $10^7$ & 1&9 & 1&5&$10^8$ & 1,3\\
Leo I............................................................................  & 250 & 8&8 & 6&      & 30&      & 1&0 $\times$ $10^7$ & 2&8 & 5&2&$10^8$ & 1,3\\
Leo II...........................................................................  & 205 & 6&7 & 4&      & 20&      & 1&0 $\times$ $10^7$ & 1&9 & 1&5&$10^8$ & 1,3\\
Sculptor.......................................................................  & 79  & 8&5 & 6&      & 30&      & 1&0 $\times$ $10^7$ & 2&8 & 5&2&$10^8$ & 1,3\\
Sextans........................................................................  & 86  & 5&8 & 2&      & 10&      & 1&0 $\times$ $10^7$ & 0&9 & 1&9&$10^7$ & 1,3\\
Ursa Major II..............................................................   & 32  & 6&7 & $\cdot$&& $\cdot$& & 3&0 $\times$ $10^8$ & 0&3 & 2&0&$10^7$ & 2,4\\
Ursa Minor..................................................................    & 66  & 15&0& 6&      & 30&      & 1&0 $\times$ $10^8$ & 2&8 & 5&2&$10^8$ & 1,3\\
%Willman I............. & 38 &&& 5&5 $\times$ $10^8$ & 0&2 & 1&10$^{15}$\\
%Segue I.................. & 23 &&&&
\end{tabular}
\end{ruledtabular}
\caption{Properties of the dwarf Spheroidals used in this study.\\
References: (1) \citet{2008ApJ...672..904P}, (2) \citet{2008ApJ...678..614S}, (3) \citet{1998ARA&A..36..435M}, (4) \citet{2007ApJ...670..313S}.}\label{tab:dsphs}
\end{table*}
\endgroup

\begin{figure*}[t]
\includegraphics[clip=]{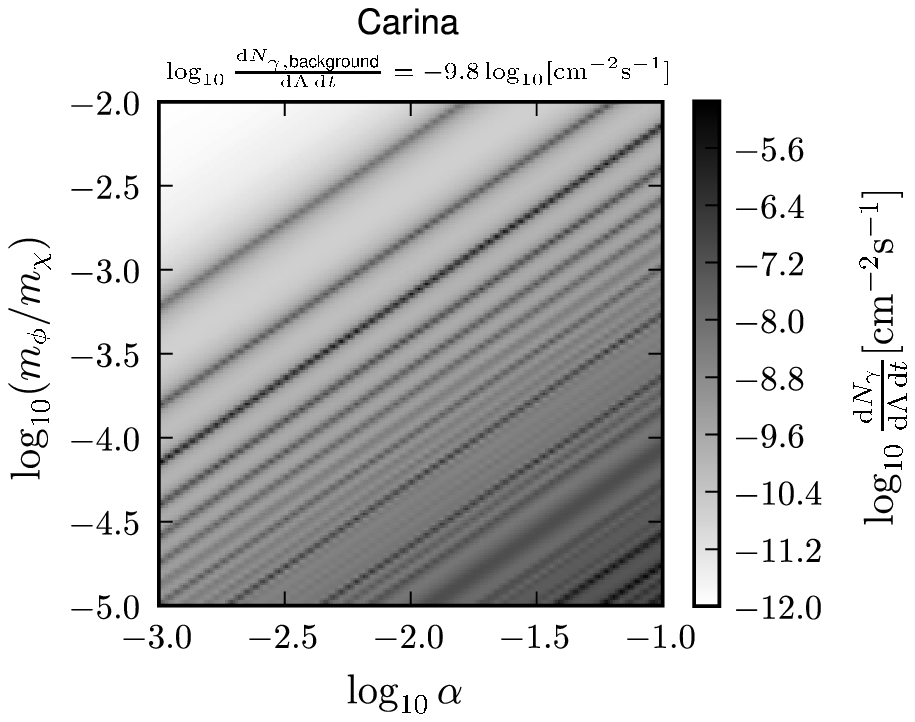}%%BoundingBox:  171 290 412 500
\includegraphics[clip=]{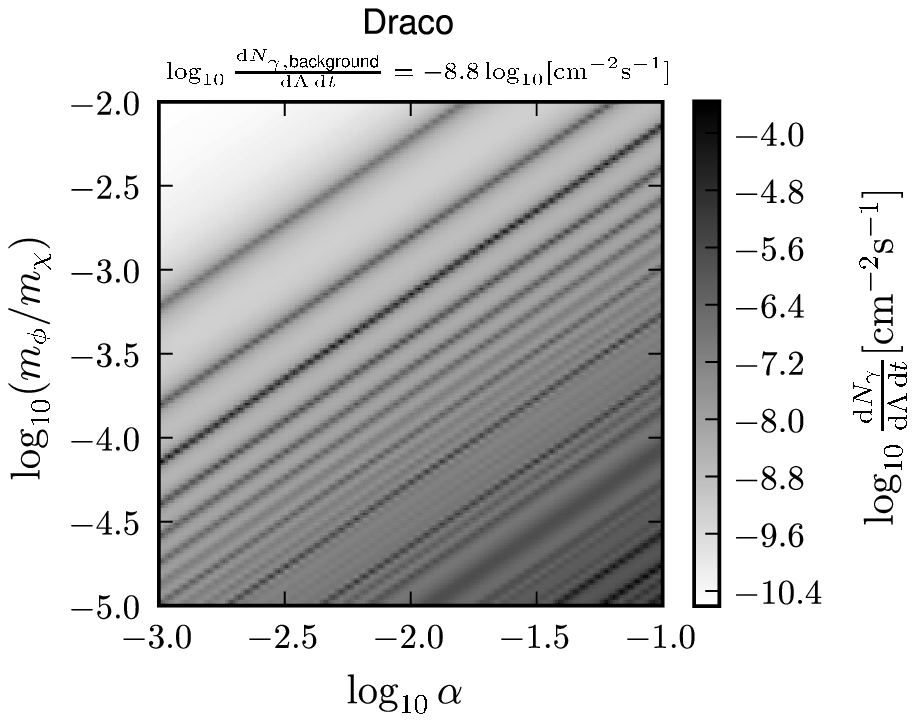}\\%%BoundingBox: 190 290 433 500
\includegraphics[clip=]{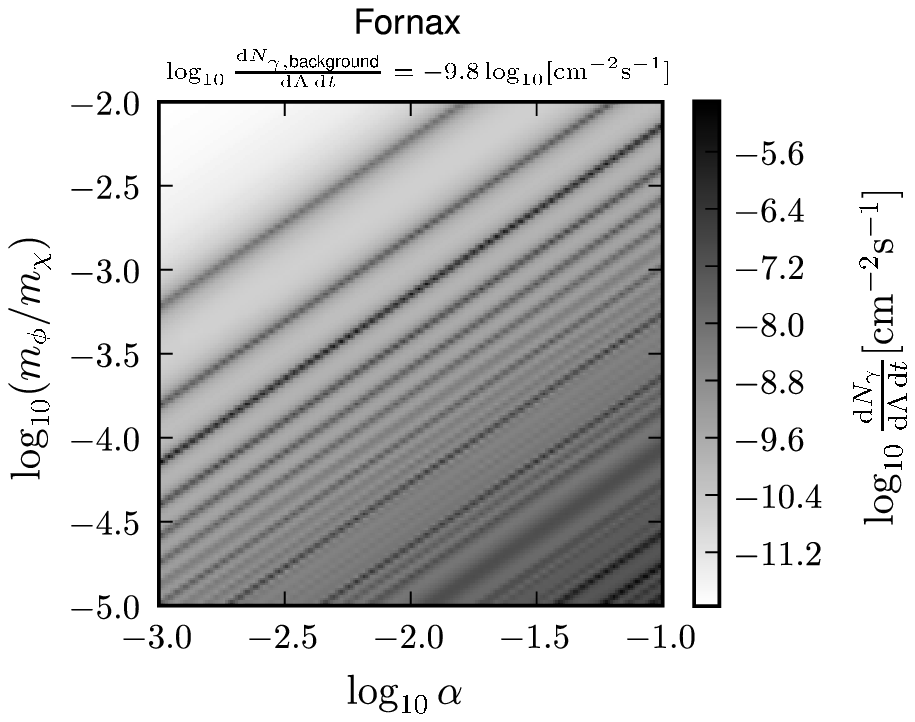}%%BoundingBox:  171 290 412 500
\includegraphics[clip=]{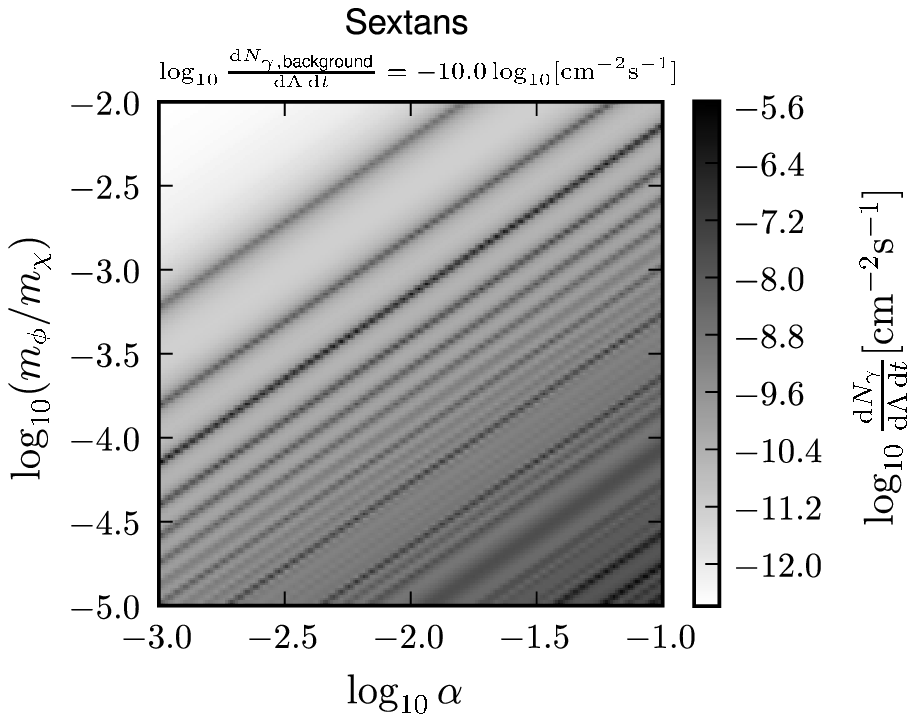}\\%%BoundingBox: 190 290 433 500
\includegraphics[clip=]{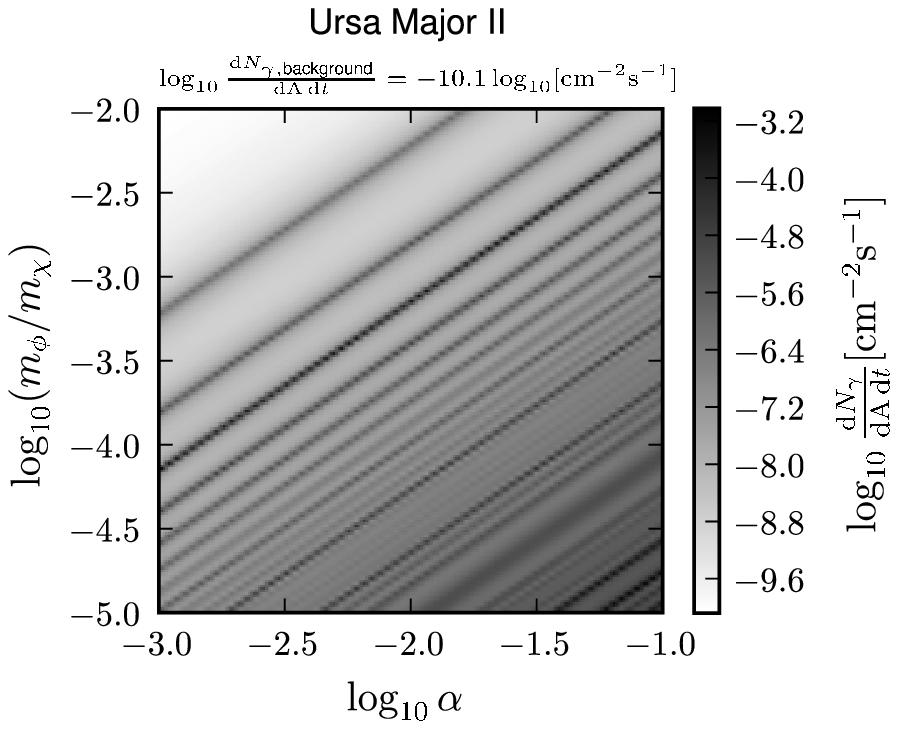}%%BoundingBox:  171 290 412 500
\includegraphics[clip=]{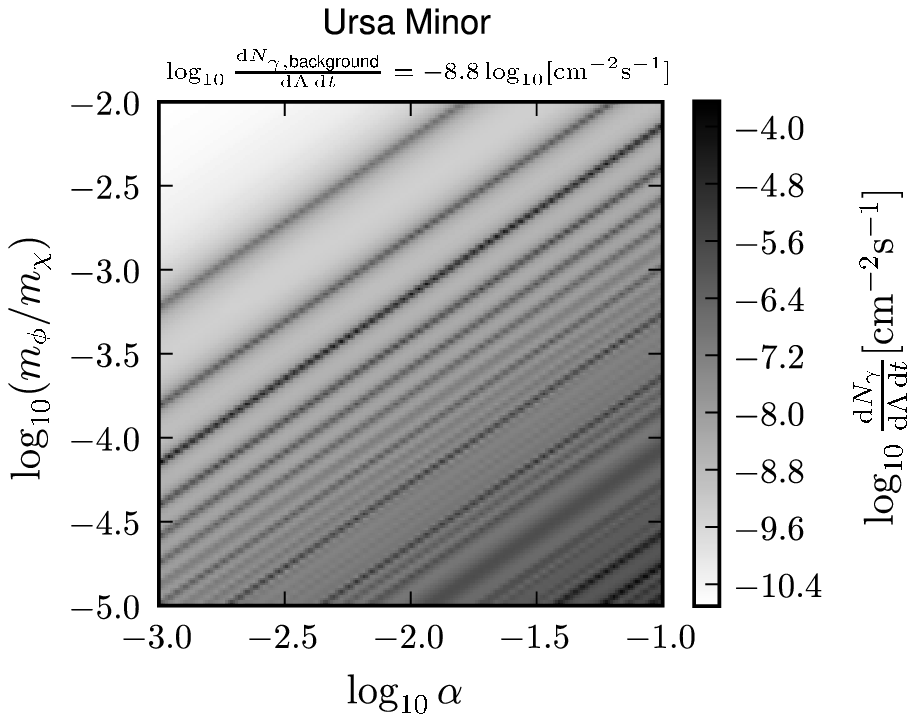}%%BoundingBox: 190 290 433 500
\caption{Predicted \DM\ annihilation fluxes for various dwarf
Spheroidal satellite galaxies. The background flux of extragalactic
photons in the relevant energy range is given under the name of the
dwarf Spheroidal. The properties of the \dSphs\ used are given in
Table \ref{tab:dsphs}.}%
\label{fig:dsph}%
\end{figure*}

In Figure \ref{fig:subboostslope} the effect of varying two more of
the many parameters introduced in the previous section is shown. The
slope of the subhalo mass function $n$ was set to -0.9 before,
corresponding to the results from the latest high-resolution numerical
simulations. In Figure \ref{fig:subboostslope} (\emph{top panels}) we
show the effect of both a shallower as well as a steeper subhalo mass
function. The value of $n = -0.7$ corresponds to the shallow slope
found in Ref.~\cite{2002PhRvD..66f3502H}. We see that the effect of
such a shallow slope is to seriously reduce the substructure boost,
which is simply a consequence of the fact that for a shallower slope
less substructure is found in smaller and more concentrated clumps. A
steeper slope such as $n = -1$, as was found in the recent \VL\
simulation \cite{2008ApJ...686..262K} leads to a substructure boost
which is about an order of magnitude larger than for the fiducial
$n=-0.9$ value.

The effect of the thermal free-streaming scale of \DM\ on the
substructure boost is shown in Figure \ref{fig:subboostslope}
(\emph{bottom panels}). This effect is relatively large since the
free-streaming scale sets the scale of the smallest and most
concentrated \DM\ halos and raising or lowering this scale therefore
changes an important part of the contribution to the substructure
boost. For very large values of the free-streaming scale such as $m_0
= 10^{-3}$ \Msun, the substructure boost becomes about an order of
magnitude smaller. When the free-streaming scale becomes very
small, much of the substructure is in very small, cold clumps, such
that the total boost becomes about an order of magnitude larger.

\section{Predicted annihilation signals for dwarf Spheroidals}\label{sec:dsphs}

Besides detection of \DM\ annihilation from the Galactic Center or the
diffuse component from the Galactic halo, there idea exists that the
best candidates for \DM\ annihilation detection might be the dwarf
Spheroidal satellite galaxies (\dSphs) of our Galaxy, since the dwarf
galaxies are believed to be the most dark matter dominated structures
in the Universe, and the \DM\ annihilation signal coming from them
would be much easier to distinguish from background sources than, \eg,
would be the case for \DM\ annihilation photons coming from the
Galactic Center. That \dSphs\ make good candidates for \DM\
annihilation detection has been much discussed before, especially for
the ``classical'', or pre-\SDSS, \dSphs\
\cite{2000PhRvD..61b3514B,2002PhRvD..66b3509T,2004PhRvD..70b3512B,2004PhRvD..69l3501E,2004PhRvD..70d3503P,2004PhRvL..93p1302H}. The
annihilation signal coming from Draco has been the subject of much
research in recent years because of its relative proximity to the
Earth
\cite{2002PhRvD..66b3509T,2004PhRvD..70d3503P,2006PhRvD..73f3510B,2006JCAP...03..003P,2007PhRvD..75b3513C,2007PhRvD..76l3509S,2007PhRvD..75h3526S,2008arXiv0812.1494P}. The
prospects for \DM\ annihilation detection from \dSphs\ has in general
been found to be rather poor, with detection with \Fermi\ being
unlikely \cite{2008arXiv0812.1494P,2009arXiv0902.4330P}. Some of the
\dSphs\ have already been the subject of observational searches for
\DM\ annihilation and some limits on the \DM\ annihilation signal have
already been set
\cite{2008ApJ...678..594W,2008ApJ...679..428A,2008APh....29...55A,2008PhRvD..78h7101D,2009ApJ...691..175A,2008arXiv0810.3561M}.

The \dSphs\ discovered by \SDSS\ have also been suggested to make good
candidates for \DM\ annihilation detection, because of their very
large mass-to-light ratios ($\sim\!10^3$)
\cite{2008ApJ...678..614S,2008arXiv0809.2781G}. In the context of the
Sommerfeld enhancement, these recently discovered \dSphs\ are
especially interesting, since they also have the smallest velocity
dispersion of all of the known satellites of our Galaxy. Predicting
the \DM\ annihilation signal from these low-surface brightness
galaxies is complicated by the dearth of kinematical tracers of the
\DM\ distribution and their contamination with background stars
\cite{2009AJ....137.3109W}. Nevertheless, detailed studies have
constrained the \DM\ density and velocity dispersion profiles of some
of these \dSphs\
\cite{2007PhRvD..75h3526S,2007ApJ...667L..53W,2008ApJ...672..904P}. We
will use some of the results on the \DM\ density and velocity profiles
found in these studies to show the extra effect of the Sommerfeld
enhancement on the predicted \DM\ annihilation signal from some of
these \dSphs.

We consider the \dSphs\ given in Table \ref{tab:dsphs} here. Eight of
the \dSphs\ are classical dwarf galaxies and we use the \DM\ density
profiles fitted to them in Ref.~\cite{2008ApJ...672..904P}. In this
the \DM\ density profiles of the \dSphs\ were not completely
determined by the available data---a degeneracy between the
parameters of the \NFW\ profile used to describe the \DM\ density and
the King profile used to fit the luminous component of the galaxies
was found---and the concentration-mass relation from \CDM\
\cite{2001MNRAS.321..559B,2001ApJ...554..114E} was used to break the
degeneracy. Since these authors expressed their results in the form of
the \rmax\ and \vmax\ parameters of the \DM\ density profiles, we use
the procedure outlined in Section \ref{sec:DMprofile} to convert these
to \rhos\ and \rs\ values. As these \vmax\ and \rmax\ parameters were
fit to an \NFW\ \DM\ profile, we will only consider an \NFW\ profile
for the \dSphs\ here, but the general conclusions about the importance
of the exact profile shape also hold in these particular cases. The
velocity dispersions of these \dSphs\ are taken from
Ref.~\cite{1998ARA&A..36..435M} and we again model these subhalos as
having a constant velocity dispersion.

We also consider some of the newly discovered \dSphs. In particular we
discuss the prospects for \DM\ annihilation detection from Coma
Berenices and Ursa Major II, and we use the best fit values to the
available data from Ref.~\cite{2008ApJ...678..614S}. Since these
authors directly quote \rhos\ and \rs\ values we do not include \rmax\
and \vmax\ values for these \dSphs. We use the velocity dispersions
determined for these galaxies in Ref.~\cite{2007ApJ...670..313S}.

In order to calculate the substructure boost to the \DM\ annihilation
cross section we need the mass of the \dSphs. Many definitions of the
mass of the galaxy could be used here, among the more appropriate ones
is the mass contained inside the tidal radius, \ie, the radius inside
of which material in the \dSph\ is safe from being tidally stripped by
the Galaxy. However, since we are considering the \DM\ annihilation
flux coming from within the radius \rminustwo\ we will use the mass
contained within this radius to calculate the substructure boost. This
is more appropriate since (1) it can be calculated straightforwardly
from the \DM\ density profile of the \dSph\ and (2) the mass contained
within \rminustwo\ is really the more relevant mass to calculate the
substructure boost, since we are only interested in contributions to
the \DM\ annihilation flux coming from within the radius \rminustwo.

Because of the similarities between some of the \DM\ density profiles
of the \dSphs\ in Table \ref{tab:dsphs} we do not show the predicted
annihilation fluxes for all of the \dSphs\ in Figure \ref{fig:dsph},
instead focusing on a representative sample of them. Inspection of
Table \ref{tab:dsphs} shows that annihilation flux of the \dSphs\ not
shown here will be similar to that of the ones shown in Figure
\ref{fig:dsph}.

At this point we can also make a connection to another quantity that
is often used to describe the astrophysical contribution to the \DM\
annihilation signal, $J$. This quantity is generally defined as
\begin{equation}
J \equiv \frac{1}{\Delta \Omega} \int_{\Delta \Omega} \dd \Omega\, \int_{\mbox{{\footnotesize los}}} \dd l\, \rho^2\, ,
\end{equation}
and is therefore equal to the line-of-sight integral over the density
squared averaged over solid angle. We can define a slight variant of
this which takes into account the Sommerfeld enhancement
\begin{equation}
\Jtilde \equiv \frac{1}{\Delta \Omega} \int_{\Delta \Omega} \dd \Omega\, \int_{\mbox{{\footnotesize los}}} \dd l\, \rho^2 \somm\, ,
\end{equation}
in which \somm\ is the Sommerfeld enhancement averaged over the
distribution of relative velocities. If we work this out for the
dwarf Spheroidals, using the expression from \eqnname\
(\ref{eq:lumrhosrsNFW}) for the line-of-sight integral averaged over
the solid angle spanned by the scale radius \rs\ we find that this
\Jtilde\ is approximately equal to the total boost \totalboost, with a
pre-factor \order(1). Therefore in the case of the \dSphs\ studied
here the \Jtilde\ can be approximately read off from the top panels of
Figure \ref{fig:subboost}.

It is clear from Figure \ref{fig:dsph} that the \DM\ annihilation flux
predicted to be coming from the \dSphs\ is very large in a large class
of models. We have to stress here that our predicted fluxes are all
for the particle physics specific parameters given in Section
\ref{sec:partphys}, \ie, \sigmaannv\ = $3 \times 10^{-26}$ cm$^3$
s$^{-1}$, \mdm\ = 700 GeV and a particular form for the \DM\
annihilation spectrum. To predict the gamma-ray flux for a particular
model with different values of these particle physics parameters one
is required to calculate the particle physics parameter specific
factor to the \DM\ annihilation flux and appropriately rescale the
values given here. In order to compare these signals to the expected
background flux we will only consider contributions to the background
flux of extragalactic gamma rays. Although there are also
contributions to the background flux from pion-decay, inverse Compton
scattering, and Bremsstrahlung, at the high Galactic latitudes at
which the dwarf Spheroidals are found these are all of the same order
or smaller than the extragalactic background
\cite{2004ApJ...613..962S}. We have integrated the extragalactic gamma
ray flux as measured by EGRET over the energy range between 4 GeV to
250 GeV and multiplied it with the solid angle that the \dSphs\ span
on the sky. The extragalactic gamma ray flux is given by a power law
\cite{1998ApJ...494..523S}
\begin{equation}
\frac{\dd \Ngamma}{\dd E \,\dd A \, \dd t\,\dd \Omega} = k\,\left(\frac{E}{E_0}\right)^{-\alpha}\, ,
\end{equation}
where $k = 7.3 \times 10^{-6}$ cm$^{-2}$ s$^{-1}$ sr$^{-1}$
GeV$^{-1}$, $\alpha$ = 2.10, and $E_0 = 0.45$ GeV. We have included
the background flux of extragalactic photons in Figure
\ref{fig:dsph}. This background flux is in all cases of the order of
10$^{-9}$ to 10$^{-10}$ photons cm$^{-2}$ s$^{-1}$, such that the \DM\
annihilation boosted by substructure and Sommerfeld enhancement
outshine this background in most of the parameter space, and by a few
orders of magnitude in most models. Given that the flux above 100 MeV
over a 55 days time-span needed to separate a point source of \DM\
annihilation from the background by \Fermi\ to give a 5$\sigma$
detection is about 10$^{-8}$ photons cm$^{-1}$ s$^{-1}$ and that the
necessary fluxes scale roughly as $t_{\mbox{{\footnotesize
exp}}}^{-1/2}$ \cite{2008JCAP...07..013B}, the \DM\ annihilation of
these \dSphs\ should be detectable by \Fermi\ in many of the models
parametrized by ($\mv/\mdm,\alpha$). The Ursa Major II \dSph\ is the
only \dSph\ for which the \DM\ annihilation dominates the background
in all the regions of parameter space. This shows that the newly
discovered, most-dark-matter-dominated satellite galaxies are possibly
the best targets for the indirect detection of \DM. We conclude that
the prospects for detecting \DM\ annihilation from the \dSphs\ with
\Fermi\ are very good in models that include a significant Sommerfeld
enhancement.

\begin{figure}
\centering
\includegraphics{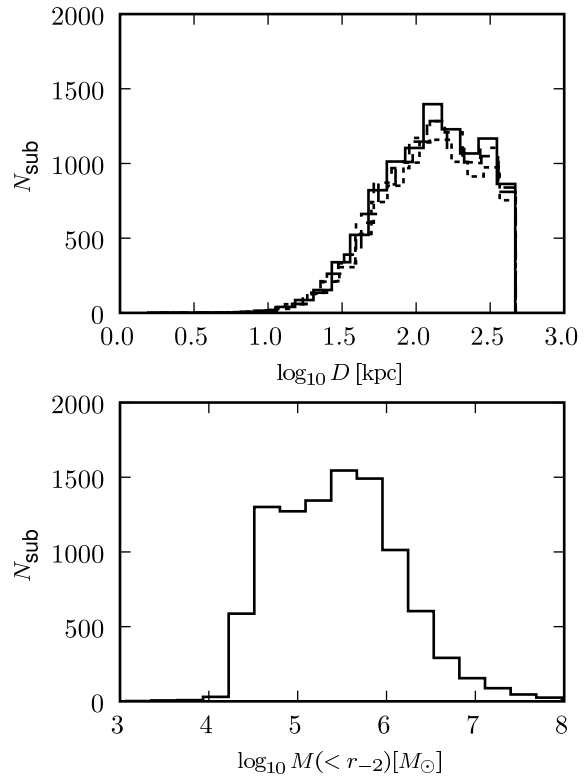}
\caption{The distances to and masses of subhalos within the tidal
radius of the main halo in \VL\ II. The distances are for three random
observers at a distance of 8.5 kpc from the galactic center of the
main halo.}%\
\label{fig:VL23_1}%
\end{figure}

\begin{figure}
\centering
\includegraphics{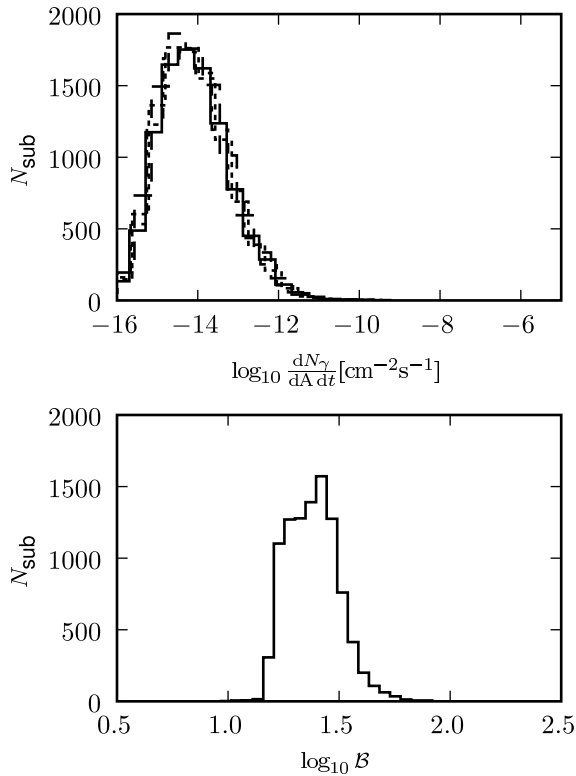}
\caption{Top: Histogram of the predicted gamma-ray flux coming from
individual subhalos contained within the tidal radius of the main halo
in \VL\ II for three random observer placed at a distance of 8.5 kpc
from the center of the main \VL\ halo for \mv/\mdm\ = 10$^{-2}$ and
$\alpha = 10^{-3}$. Bottom: Histogram of the total boosts of the
subhalos for the same values of \mv/\mdm\ and $\alpha$.}%\
\label{fig:VL23_2}%
\end{figure}

\begin{figure}
\centering
\includegraphics{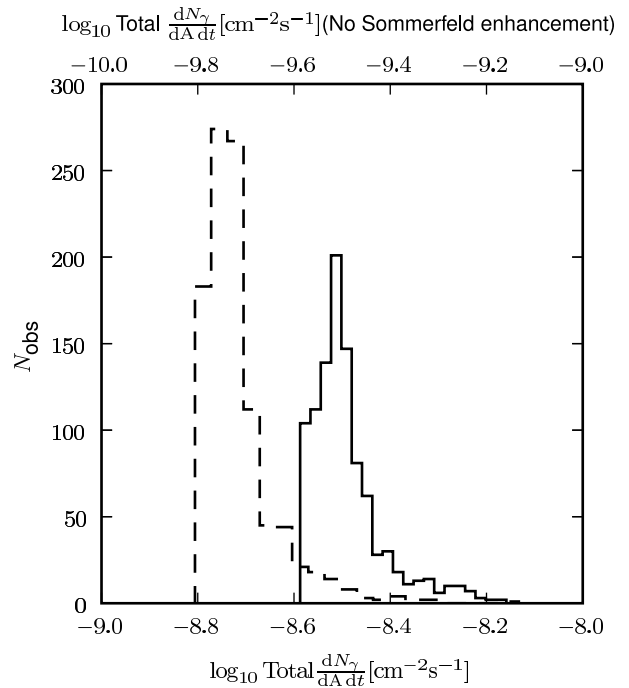}
\caption{Total gamma-ray flux from subhalos in \VL\ II for 1,000
random observers at 8.5 kpc from the center of the main \VL\ halo for
\mv/\mdm\ = 10$^{-2}$ and $\alpha = 10^{-3}$. For comparison we show
the total gamma-ray flux from subhalos in the absence of Sommerfeld
enhancement as the dashed histogram and the top horizontal axis.}%\
\label{fig:VL23_3}%
\end{figure}

\begin{figure}
\centering
\includegraphics{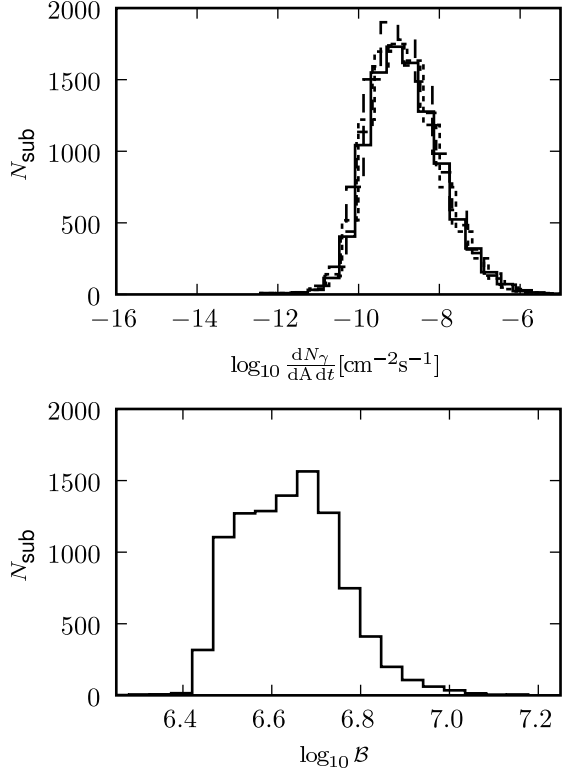}
\caption{Same as Figure \ref{fig:VL23_2} but for \mv/\mdm\ = 10$^{-5}$
and $\alpha = 10^{-1}$.}%
\label{fig:VL51_1}%
\end{figure}

\begin{figure}
\centering
\includegraphics{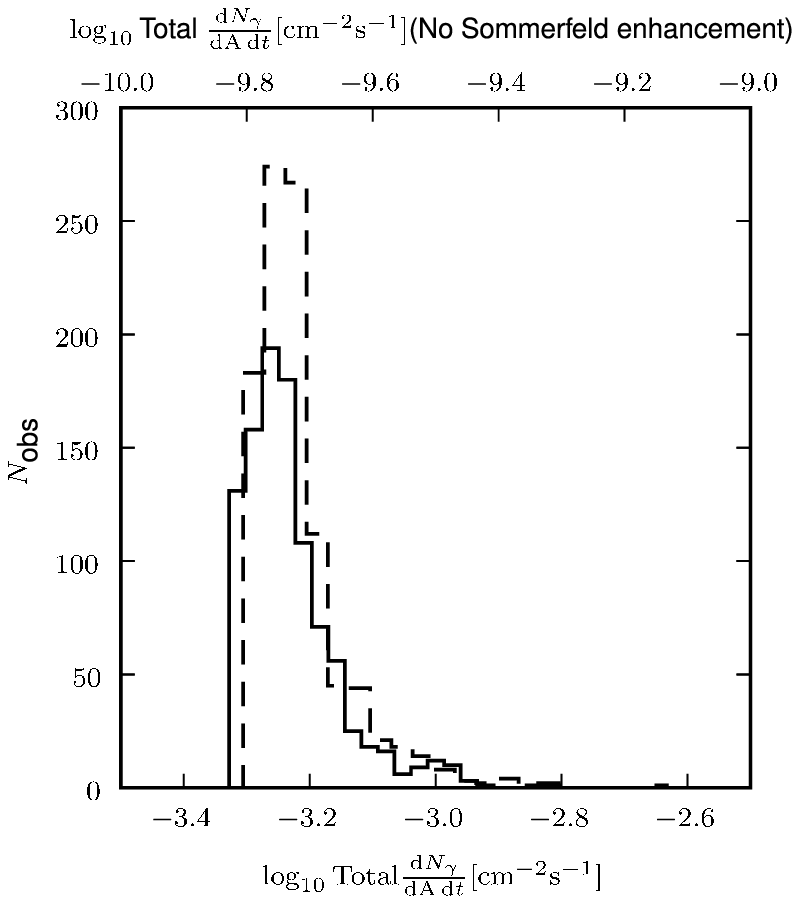}
\caption{Same as Figure \ref{fig:VL23_3} but for \mv/\mdm\ = 10$^{-5}$
and $\alpha = 10^{-1}$.}%
\label{fig:VL51_2}%
\end{figure}

\section{The annihilation signal from subhalos in \VL\ II}

Besides the satellite galaxies that have been observed in the Galactic
halo \CDM\ predicts that the halo is filled with hundreds of
individual subhalos---a fact which gives rise to the so-called
``missing satellites problem''
\cite{1999ApJ...522...82K,1999ApJ...524L..19M}. Each of these subhalos
has a mass of the order of the masses of the observed \dSphs\ or
smaller, and thus will be kinematically cold with respect to the main
halo. Thus, the total \DM\ annihilation signal coming from all of the
substructure will be boosted by the Sommerfeld enhancement described
above, such that an appreciable \DM\ annihilation signal could be
expected from all of the substructure. In this section we get a
feeling for the expected signal from \DM\ substructures in a typical
Galaxy-sized halo. If the \DM\ annihilation signal from subhalos
without star formation could be detected this could give a clear
signal that the \DM\ exists and would open up a whole new arena for
testing \CDM.

\begin{figure}
\centering
\includegraphics{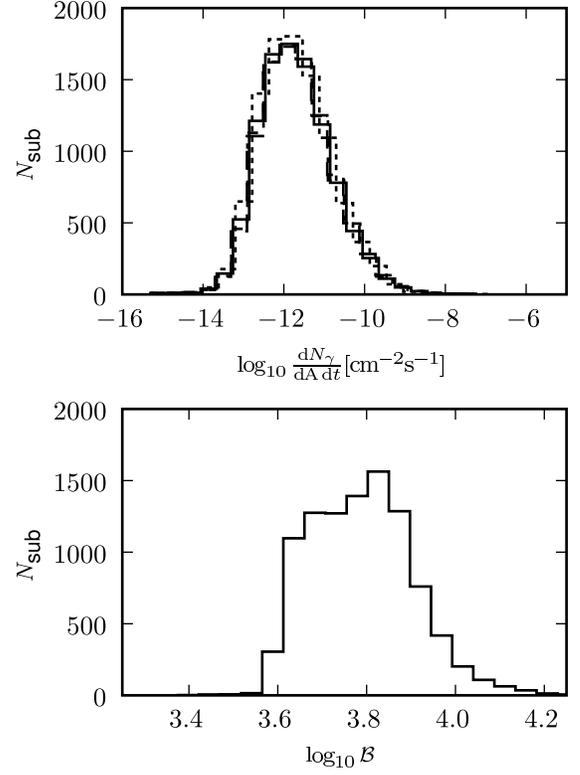}
\caption{Same as Figure \ref{fig:VL23_2} but for a resonance:
\mv/\mdm\ = 10$^{-3.84375}$ and $\alpha = 10^{-2.625}$.}%
\label{fig:VLresonance1}%
\end{figure}

\begin{figure}
\centering
\includegraphics{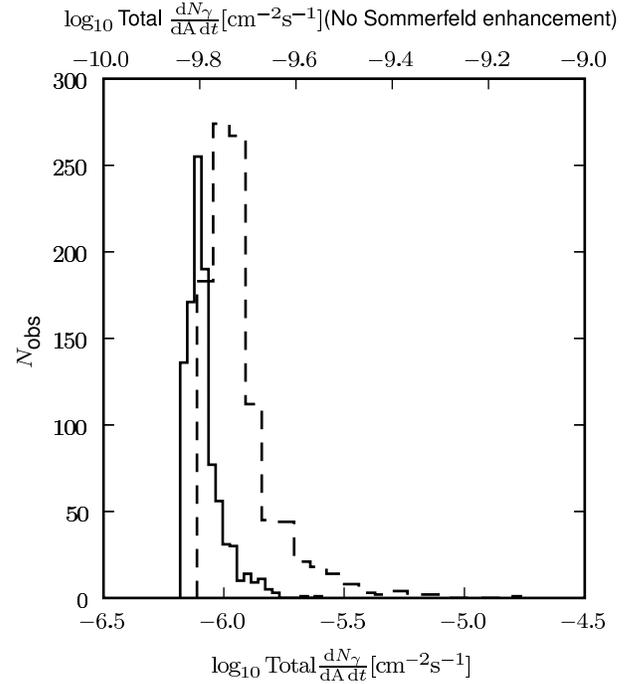}
\caption{Same as Figure \ref{fig:VL23_3} but for a resonance:
\mv/\mdm\ = 10$^{-3.84375}$ and $\alpha = 10^{-2.625}$.}%
\label{fig:VLresonance2}%
\end{figure}

We use the distribution of \DM\ subhalos from the \VL\ II numerical
$N$-body simulation\footnote{Available at
\texttt{http://www.ucolick.org/\urltilda$\!$diemand/vl/data.html}}
\cite{2008Natur.454..735D}. \VL\ II is a state-of-the-art $N$-body
simulation of a Galaxy-sized halo since redshift 104.3. It sampled the
region of the main halo with 1.1 $\times$ 10$^9$ particles, each with
a mass of 4,100 \Msun. Over 40,000 subhalos can be resolved within 402
kpc of the center of the main halo and 20,047 subhalos are detected
which had a peak circular velocity larger than 4 km s$^{-1}$ at some
time. The properties of the subhalos in this latter sample can be
accurately established, \ie, they are unaffected by resolution
effects. We select from this sample the 9,830 subhalos which lie
within the tidal radius of the main halo, which is approximately 462
kpc. We show the distribution of these subhalos in mass and distance
from the Sun in Figure \ref{fig:VL23_1}. We have placed the Sun at
random locations at a distance of 8.5 kpc of the center of the main
halo, corresponding to an earth-based-observer-like vantage point. The
distribution of the distances in Figure \ref{fig:VL23_1} is given for
three of these random observers and the distribution is very similar
for all three observers. We see that most of the subhalos have masses
in the range 10$^4$ to 10$^6$ \Msun, with subhalos with masses of the
order of the \dSphs\ being relatively rare. The lower limit is of
course set by the finite numerical resolution of the simulation and
doesn't correspond to a physical cut-off---indeed, the physical
cut-off lies at the thermal free-streaming scale. The distribution of
the distance to the subhalos shows that the distribution of the
subhalos is strongly radially anti-biased, with a sharp cut-off caused
by our cut at the tidal radius of the main halo. This radial anti-bias
is a consequence of the fact that there is much more space far away
from the center of the galaxy, and because of the strong tidal forces
near the center of the galaxy, which strongly affects halos which pass
near the center of the galaxy.

For each halo we can estimate the velocity dispersion using the
mass-velocity dispersion relation given in \eqnname\
(\ref{eq:vdispscaling}) and we can calculate the characteristic
density \rhos\ and the characteristic length-scale \rs\ from the
values of \vmax\ and \rmax\ given for each subhalo, by the procedure
outlined at the end of Sec.~\ref{sec:DMprofile}. For a given set of
particle physics parameters \mv/\mdm\ and $\alpha$ this allows us to
calculate the total boost to the \DM\ annihilation cross section for
each of the subhalos, and by using the assumptions about the particle
physics given in Sec.~\ref{sec:partphys} we can also calculate the
photon flux from \DM\ annihilation coming from each subhalo, and the
total gamma-ray flux coming from all of the subhalos combined.

We only cover some typical cases for the Sommerfeld enhancement here
which suffices to show the general trend. We consider the particle
physics parameters \mv/\mdm\ and $\alpha$ which give rise to the
smallest Sommerfeld enhancements in Figures \ref{fig:VL23_2} and
\ref{fig:VL23_3}, \ie, \mv/\mdm\ = 10$^{-2}$ and $\alpha$ = 0.001. The
top panel of Figure \ref{fig:VL23_2} shows the distribution of
gamma-ray fluxes from \DM\ annihilation for the subhalos, while the
bottom panel shows the distribution of the total boost factors. As can
be seen, the total boost factors are quite small: the typical boost
factor is only about 30 with the highest boost factors being about
100. The total gamma-ray flux coming from subhalos is then also quite
small in this case, as is shown in Figure \ref{fig:VL23_3}, in which
the total gamma-ray flux is shown for 1,000 random observers at a
distance of 8.5 kpc from the galactic center. At most observation
locations this total flux is about 10$^{-7}$ cm$^2$ s$^{-1}$, with the
high-end tail not extending further than about half an order of
magnitude from this. For comparison the dashed histogram in Figure
\ref{fig:VL23_3} shows the same distribution of total gamma-ray flux
for different random observers in the absence of any Sommerfeld
enhancement. This confirms that the total gamma-ray flux with
Sommerfeld enhancement is about an order of magnitude larger than
without Sommerfeld enhancement. 

Figures \ref{fig:VL51_1} and \ref{fig:VL51_2} show the same for a
different extremum of the Sommerfeld enhancement: the case in which
\mv/\mdm\ = 10$^{-5}$ and $\alpha$ = 0.1. As we saw before, in this
case the Sommerfeld enhancement can reach very high values, and thus
we see that the typical \DM\ annihilation signal coming from \DM\
subhalos is large: the typical flux for one subhalo is equal to the
total flux in the scenario described in the previous paragraph. The
total boost factors for most of the halos are very large, of the order
of 10$^7$, which gives rise to this very large photon-flux. The total
predicted flux in this case is many orders of magnitude larger than
the total flux in the other extremum of the Sommerfeld enhancement and
about seven orders of magnitude larger than the total gamma-ray flux
in the absence of any Sommerfeld enhancement. Thus, the exact value of
the parameters \mv/\mdm\ and $\alpha$, upon which the Sommerfeld
enhancement depends, matters a great deal for the predicted gamma-ray
flux.

Finally, we look at the expected signal coming from subhalos when the
Sommerfeld enhancement is at a resonance, the same distribution as
before are shown in Figures \ref{fig:VLresonance1} and
\ref{fig:VLresonance2}. Again we find large boost factors and large
individual and total gamma-ray fluxes, although they are smaller than
the values found for the \mv/\mdm\ = 10$^{-5}$ and $\alpha$ = 0.1
parameter set. However, since these values are near a resonance in a
region of parameter space around which the predicted signal looks more
like that for the parameter set \mv/\mdm\ = 10$^{-2}$ and $\alpha$ =
0.001, this again shows that whether dark halos, \ie, subhalos without
star formation, can be detected through \DM\ annihilation products
depends very sensitively on the particle physics parameters upon which
the Sommerfeld enhancement depends.

In conclusion we can say that in a scenario in which there is
Sommerfeld enhancement there is much more hope to detect dark subhalos
than in more conventional scenarios, but that the details of the
prediction depend strongly on the underlying particle physics.

\section{Conclusion}

The recently proposed Sommerfeld enhancement to the \DM\ annihilation
cross section forces us to reconsider statements made in the past
about the expected signal of \DM\ annihilation from cold substructures
in the halo of the Galaxy. Since the Sommerfeld enhancement depends
critically on the velocity dispersion of the substructure in
question, as well as on the details of the particle physics model used
to describe the attractive (or, possibly, repulsive) force that gives
rise to the Sommerfeld enhancement, the contributions from the
specific particle physics model and the details of the \DM\
distribution in the substructure can no longer be as cleanly separated
as they have been before. This means that in a scenario in which there
is a significant Sommerfeld enhancement, previous claims of ``particle
physics independent'' conclusions about the \DM\ annihilation signal
from substructure (\eg,
\cite{2007PhRvD..75h3526S,2008Natur.456...73S}) need to be
reconsidered in this new framework. This complicates the analysis but
it opens up new possibilities for detecting \DM\ annihilation and
learning about the properties of the dark sector. As we showed above,
the combination of the Sommerfeld enhancement and the substructure
that is believed to exist down to small scales leads to very large
boost factors because of the combined effect of the higher densities
of small subhalos as well as the fact that they are kinematically
colder than the larger halos. This leads to predictions for the \DM\
annihilation gamma-ray flux that are much larger than was previously
believed. As we showed in Sec.~\ref{sec:dsphs}, in models in which
there is a significant Sommerfeld enhancement at the level of the
Galactic halo and its subhalos, the \DM\ annihilation signal from the
dwarf Spheroidals outshines the extragalactic background flux and the
predicted gamma-ray fluxes are so large that they should be detectable
by \Fermi. Given a detailed knowledge of the density structure and
velocity dispersion profile of the dwarf Spheroidal galaxies, much can
then be learned if we are able to detect the \DM\ annihilation signal
coming from these \dSphs, by considering the relative signal and using
the results of Sec.~\ref{sec:dsphs}, or by using the detailed velocity
dispersion profile of the \dSphs\ \cite{2009arXiv0902.0362R}.

We have focused on the consequences of the Sommerfeld enhancement for
\DM\ subhalos in a \LCDM\ cosmology, using the latest results from
high-resolution numerical $N$-body simulations to inform our
assumptions about the abundance and properties of the
substructure. Under these assumptions we were able to show that there
are very large boosts to the \DM\ annihilation cross section from the
combined effect of substructure boosts and Sommerfeld enhancement in
large parts of the parameter space consisting of the parameters of the
attractive Yukawa type force between the \DM\ particles which is
responsible for the Sommerfeld enhancement. These regions of parameter
space are in no way unrealistic. Indeed, in one of the most
straightforward interpretations of the ATIC/PAMELA results, the sharp
cut-off around 1 TeV seen in the ATIC spectrum is interpreted as being
a consequence of the fact that the dark matter mass is about 700
GeV. The absence of an excess in anti-protons similar to the excess in
positrons seen by PAMELA can be explained by positing that the dark
matter particles annihilate to an intermediate boson $\phi$, which is
lighter than a proton (or even, lighter than the effective quark mass
$\Lambda_{\mbox{\footnotesize QCD}}$) in order to kinematically forbid
the annihilation into protons \cite{2008arXiv0811.3641C}. This
naturally leads to a \mv/\mdm $\lesssim$ $10^{-3}$. The coupling
constant of the new force in the dark sector is also generally $\alpha
\gtrsim 10^{-2}$ \cite{ArkaniHamed:2008qn,2008arXiv0812.0360L}. Thus,
from Figure \ref{fig:subboost}, we expect there to be a rather large
boost to the \DM\ annihilation cross section in these models. This
leads to particularly large \DM\ annihilation signals predicted for
the \dSphs\ (see Figure \ref{fig:dsph}).

We have not discussed the substructure boost to the \DM\ annihilation
signal from unresolved substructure, such as very small subhalos and
\DM\ tidal debris from disrupted subhalos, in the Galactic halo. This
is, however, interesting in the light of the Sommerfeld enhancement,
since it might lead to a different prediction about the relative
magnitudes of the annihilation signal coming from the resolved
Galactic halo and unresolved halo, as well from the resolved
substructures in the Galactic halos. It is already clear that in the
generic 1/$v$ Sommerfeld enhancement the main halo, with its typical
velocity dispersion of $\sim\!150$ km s$^{-1}$, will receive a much
smaller enhancement than the \dSphs, with their typical velocity
dispersions of 10 km s$^{-1}$. This, however, ignores the effects of
saturation and resonances, which might be important in the specific
particle physics model adopted for the Sommerfeld
enhancement. Additionally, while the velocity dispersion of the
Galactic halo might be of the order of 100 km s$^{-1}$, the actual
velocity dispersion relevant to the Sommerfeld enhancement might be
much smaller for halo \DM\ particles. If a large fraction of the \DM\
in the Galactic halo is part of a cold stream---and various numerical
simulations predict that large parts of the halo, if not all of the
halo, are made up of such cold streams with typical velocity
dispersions of the order of 1 km s$^{-1}$
\cite{1999MNRAS.307..495H,2003MNRAS.339..834H}---then, in addition to
any boost caused by density enhancements in the stream
\cite{2008arXiv0811.1582A}, the boost from the Sommerfeld enhancement
could also play a large role, as the relevant relative velocity of
\DM\ particles is set by the velocity dispersion of particles in the
stream, and not by the velocity dispersion of the streams in the halo,
which is what the 100 km s$^{-1}$ Galactic halo velocity dispersion
corresponds to in these scenarios. Thus, the expected \DM\
annihilation signal from the main halo could be much larger than that
naively expected for a halo with a velocity dispersion of 100 km
s$^{-1}$.

{\bf Acknowledgments:} It is a pleasure to thank Ilias Cholis, Mulin
Ding, David Hogg, and Neal Weiner for comments and assistance. This
research was partially supported by NASA (grant NNX08AJ48G).

\end{document}